\documentclass[aps,prx,reprint,groupedaddress]{revtex4-2}
\usepackage{graphicx}% Include figure files
\usepackage{amssymb}
\usepackage[T1]{fontenc}
\usepackage{float}
\usepackage{dcolumn}% Align table columns on decimal point
\usepackage{physics}
\usepackage{bm}% bold math
\usepackage{bbold}
\usepackage{tikz}
\usetikzlibrary{matrix, positioning}
\graphicspath{ {figs/} }
\usepackage{comment}
\usepackage[english]{babel}
\babelprovide{en}

\newcommand{\be}{\begin{equation}}
	\newcommand{\ee}{\end{equation}}
\newcommand{\bea}{\begin{eqnarray}}
	\newcommand{\eea}{\end{eqnarray}}
\newcommand{\Eq}[1]{Eq.\,(\ref{#1})}% \Eq{abc}
\newcommand{\Eqs}[2]{Eqs.\,(\ref{#1}) and (\ref{#2})}

\newcommand{\Eqss}[2]{Eqs.\,(\ref{#1})--(\ref{#2})}
\newcommand{\Fig}[1]{Fig.\,\ref{#1}}% \Fig{fig:abc}
\newcommand{\Figs}[2]{Figs.\,\ref{#1} and \ref{#2}}

\newcommand{\Sec}[1]{Sec.\,\ref{#1}}% \Sec{sec:abc} sic!byc konsekewntnym \label{sec:xx} \Sec{sec:xx}
\newcommand{\App}[1]{Appendix\,\ref{#1}}% \Sec{sec:abc}

\newcommand{\rmand}{\quad{\rm and}\quad}

\begin{document}
	
\title{Optimization of path-integral tensor-multiplication schemes for open quantum systems}

\author{L. M. J. Hall\textsuperscript{1}}
\email{Luke.Hall415@gmail.com}
\author{A. Gisdakis\textsuperscript{2}}
\author{E. A. Muljarov\textsuperscript{1}}
\email{Egor.Muljarov@astro.cf.ac.uk}
\affiliation{
		\textsuperscript{1}School of Physics and Astronomy, Cardiff University, Cardiff CF24 3AA, United Kingdom, \\
		\textsuperscript{2}School of Physics and Astronomy, University of Birmingham, Edgbaston, Birmingham B15 2TT, United Kingdom
}
\date{\today}

\begin{abstract}
Path-integral techniques are a powerful tool used in open quantum systems to provide an exact solution for the non-Markovian dynamics. However, the exponential  scaling of the tensor size with quantum memory length of these techniques limits the applicability when applied to systems with long memory times. Here we provide a general optimization of tensor multiplication schemes for systems with pair correlations and finite memory times. This optimization effectively reduces the tensor sizes by using a matrix representation of tensors combined with singular value decomposition to filter out negligible contributions. This approach dramatically reduces both computational time and memory usage of the traditional tensor-multiplication schemes.
While more memory-efficient representations exist, this approach enables a consistent extrapolation scheme for the rapid estimation of the exact value. As a demonstration, we apply it to the Trotter decomposition with linked cluster expansion technique, and use it to investigate a quantum dot-microcavity system at large coupling strengths. We also apply the optimization in a case where the memory time is very long -- specifically in a system containing two spatially separated quantum dots in a common phonon bath.
\end{abstract}

\maketitle

\section{Introduction}
%both cumulant expansion (in IB model) and influence functional approach in Feynman path integral
%use the fact that due to the harmonic non interacting bath, a gaussian form emerges
%and the gaussian statistics means higher order terms ca n be expressed in terms of 2 time correlations
%this is because of the factorizability of products into two time correlations via Wick's theorem
%both techniques package an infinite series of correlations into a compact exponential form, relying on
%the gaussian nature of harmonic bath states.
Decoherence phenomena and energy relaxation dynamics of open quantum systems are characterized by the interaction between the system and its surrounding environment (the bath). In the simplest case of a weak interaction with the environment, it can be assumed that the bath lacks memory (i.e., is Markovian) and remains uncorrelated with the system. This assumption allows the use of Born and Markov approximations~\cite{breuerTheoryOpenQuantum2002, devegaDynamicsNonMarkovianOpen2017, breuerColloquiumNonMarkovianDynamics2016} resulting in time-local equations of motion. This is also valid if the effect of the environment on the system occurs on a much larger timescale than the correlation time of the environment.
However, many quantum systems deviate from this idealized case, where memory effects play a critical role and render the Born-Markov approximation invalid. In such non-Markovian regimes, the system’s evolution depends on its past interactions, leading to complex phenomena~\cite{krummheuerTheoryPureDephasing2002, hohenesterPhononassistedTransitionsQuantum2009, stockAcousticOpticalPhonon2011, reiterDistinctiveCharacteristicsCarrierphonon2019}. Accurately capturing these non-Markovian effects is essential but comes with significant computational challenges, often limiting the scope of treatable systems and coupling regimes.

The typical approaches to solving the dynamics in non-Markovian open quantum systems can be divided into approximate and numerically-exact methods. For example, in a quantum dot (QD)-cavity system coupled to a bath of acoustic phonons, the approximate treatments are typically limited to specific parameter regimes, such as strong QD-cavity coupling and weak electron-phonon interaction~\cite{kaerNonMarkovianModelPhotonAssisted2010}, and may be improved by perturbative and variational approaches to the phonons \cite{nazirModellingExcitonPhonon2016}.
However, for stronger electron-phonon interaction, phonons play a more significant role that may require treating the system in a numerically exact way~\cite{hughesInfluenceElectronacousticPhonon2011,MorreauArX20}.

In contrast, numerically exact techniques, such as Feynman's path integral formulation, are very well suited for system-bath dynamics as they avoid dealing with the large Hilbert space of the bath by targeting the system's reduced density matrix. The formulation takes into account the effects of a harmonic bath on the system dynamics through the well known Feynman-Vernon influence functional \cite{feynmanQuantumMechanicsPath2010}, valid for any system-bath coupling strength. In practice, the issue is that the influence functional is nonlocal in time, meaning that the coordinates of a path at any particular time point are connected to those at all other time points, leading to full temporal entanglement. As a result, the number of correlations that must be accounted for grows with each time step, leading to an exponential scaling with propagation time and thus restricting simulations to short timescales. However, it is known that these nonlocal correlations decay over time, and the influence functional operates within a finite time interval called the memory time. This fact forms the basis which resulted in the development of the iterative quasi-adiabatic propagator path integral approach (i-QuAPI), which exploits the finite memory time to evaluate the dynamics of the reduced density matrix for an arbitrary time length, not limited to only short times \cite{ makarovPathIntegralsDissipative1994,makriTensorPropagatorIterative1995a,makriTensorPropagatorIterative1995,makriNumericalPathIntegral1995,shaoIterativePathIntegral2001,shaoIterativePathIntegral2002}. By combining Trotter's decomposition \cite{trotter1959product,suzuki1976generalized} with the finite memory time, the latter is discretized into a finite number of time steps, and only the information at each of the time points due to the discretization must be stored to simulate the quantum dynamics.
The number of these time steps within the memory time is commonly referred to as the number of neighbors $L$ \cite{morreauPhononinducedDephasingQuantumdotcavity2019} determining the number of correlations which form the influence functional. Thus, the finite memory time results in an influence functional of a fixed size, leading to linear scaling of the computational time with the number of propagations.
Several early works have used numerically exact path-integral techniques to simulate dynamics, including the simulation of exciton dynamics in photosynthetic complexes \cite{nalbachIterativePathintegralAlgorithm2011}. Furthermore, circuit quantum electrodynamics was explored using exact path integrals in the early 2000s \cite{goorden_entanglement_2004,thorwart_dynamics_2004}. Similarly, acoustic phonons in double quantum dots were treated numerically exactly \cite{thorwart_non-markovian_2005}.

%To clarify, the number of time steps within the memory time depends on the discretization. If there is a finer discretization, there will be more steps within the fixed memory time, and thus, more correlations within the influence functional. The size of the influence functional is what grows exponentially with the number of time steps, or neighbors, contained within the memory time --- however once this is constructed, there is linear scaling with time when propagating the system.
However, the memory storage requirements can quickly become too large in these path-integral approaches if one wishes to have a finer discretization, and in some systems convergence is not achievable. To address this, filtering techniques ~\cite{simFilteredPropagatorFunctional1997, simQuantumDynamicsSystem2001, lambertMemoryPropagatorMatrix2012}, modified truncation schemes \cite{liuImprovedMemoryTruncation2024}, and in some regimes blip decomposition ~\cite{makriBlipDecompositionPath2014,makriIterativeBlipsummedPath2017} have been developed to offer improvements to the storage requirements or extend the applicability of the existing path-integral schemes to longer memory times. However, there is still great difficulty in accurately modeling systems where memory effects from the environment are significant, such as energy transfer processes with long coherence times (e.g., photosynthetic complexes \cite{nalbach_exciton_2011}) or multi-qubit decoherence in structured environments \cite{gribbenExactQuantumDynamics2020,hallControllingDephasingCoupled2025}.
Beyond the increased computational demands associated with finer time discretization, path-integral methods face an exponential memory requirement as the system size increases, particularly in many-body or multi-level configurations. For QD-cavity problems, recent optimizations have mitigated the exponential scaling associated with the number of system levels by exploiting symmetries in the environment coupling \cite{cygorek_nonlinear_2017}. More recently, developments to QuAPI have largely resolved the memory footprint issues for systems with many states \cite{makriSmallMatrixPath2020}.

There are various other path-integral approaches which all exploit the finite memory time of the bath in an attempt to simulate the quantum dynamics \cite{barthPathIntegralDescription2016,richterCoarsegrainedRepresentationQuasi2017,morreauPhononinducedDephasingQuantumdotcavity2019,linkOpenQuantumSystem2024}. Of recent interest, an approach utilizing modern tensor network (TN) techniques called the time-evolving matrix product operator (TEMPO) algorithm \cite{strathearnEfficientNonMarkovianQuantum2018} (later packaged as OQuPY \cite{fuxOQuPyPythonPackage2024}) has provided exceptional reductions in overall memory requirements. Several other works build on the TEMPO algorithm, such as including an off-diagonal system bath coupling to the Hamiltonian, leading to multi-time correlations \cite{richterEnhancedTEMPOAlgorithm2022}. Or refinements exploiting the symmetries of the influence functional, improving efficiency of the numerical simulations \cite{jorgensenExploitingCausalTensor2019,LinkPRL24}. However, while TEMPO can achieve large values of neighbors $L$, the singular value decomposition (SVD) threshold must be decreased as $L$ increases, to ensure an accurate calculation, which in turn increases computational resource demands. An automated compression of environments (ACE) algorithm \cite{cygorekACEGeneralpurposeNonMarkovian2024a} is another tensor network approach, similar to TEMPO, which further reduces requirements by concentrating only on the most relevant degrees of freedom of the bath.

In this work, we focus on one of the aforementioned, numerically exact tensor multiplication scheme, developed in parallel with TEMPO, which is the Trotter decomposition technique with linked cluster expansion~\cite{morreauPhononinducedDephasingQuantumdotcavity2019}. This  method has already been applied to various systems and calculations of several observables or elements of the reduced density matrix, such as four-wave mixing polarization in QD-cavity systems~\cite{sirkinaImpactPhononEnvironment2023a}, linear polarization in multi-qubit systems~\cite{hallControllingDephasingCoupled2025}, and population dynamics in F\"{o}rster coupled QDs~\cite{sirkina_forster_2025}. However, being a tensor multiplication scheme, it also suffers from the exponential scaling with time steps included in the finite memory time.

In this paper, we develop an intuitive optimization scheme to the propagator and influence functional tensors used in a general path-integral based tensor multiplication approach with pair correlations and finite memory time. For demonstration, we use the Trotter decomposition technique with linked cluster expansion as a reliable tool for calculation of the exact quantum dynamics of an open quantum system.
To ease the presentation, we introduce the optimization scheme first in a simpler case of a linear optical response of a QD-cavity system treated in Ref.\,\cite{morreauPhononinducedDephasingQuantumdotcavity2019}, allowing us to clearly demonstrate and intuitively explain the optimization scheme. The technique developed in Ref.\,\cite{morreauPhononinducedDephasingQuantumdotcavity2019} is referred to as the reference method.

The optimization is applicable to systems of any basis size and to any elements of the reduced density matrix, and has been applied in Liouville space to determine the population dynamics in F\"orster coupled QDs \cite{sirkina_forster_2025}. The optimization reduces the influence functional tensor sizes by remapping them to normal matrices and utilizes SVD to filter out negligible contributions, in this way reducing storage requirements and computational time. Furthermore, we express the propagator tensor as a product of exponentials each describing pair correlations and provide an efficient way to propagate the system without storing the tensor at all. The reduction in storage requirements results in approximately twice the maximum number of neighbors $L$ (discrete time steps in the influence functional) included than previously available in the reference method. This provides better accuracy in systems with longer memory times and access to new parameter regimes. For reference, to achieve the new level of accuracy using the reference method for the calculations performed in this work, it would require over 50 million GB of RAM, but now this is available on a standard PC.

The main conceptual difference of our optimization scheme compared to other approaches, e.g. \cite{strathearnEfficientNonMarkovianQuantum2018, fuxOQuPyPythonPackage2024,  richterEnhancedTEMPOAlgorithm2022, jorgensenExploitingCausalTensor2019, LinkPRL24, cygorekACEGeneralpurposeNonMarkovian2024a}, is that we do not employ TN techniques. The TN representation directly provides the first level of optimization in terms of memory requirements, but consequently introduces many extra degrees of freedom and links which need to be controlled and optimized further. This can lead to some difficulties determining convergence and also largely incorrect results in some cases if not careful with the chosen parameters. Although TNs provide a more memory-efficient representation for the tensors, this approach enables the usage of a consistent and easy to use extrapolation scheme. The extrapolation procedure accurately approximates the exact ($L\rightarrow \infty$) long-time dynamics, allowing for the extraction of parameters such as the dephasing rates.
% The reason the extrapolation does not work in other techniques is that it requires calculations across a range of neighbours, and the {\it same} SVD threshold must be maintained across all calculations for consistency and smooth convergence behaviour. However, in other techniques, as the number of neighbors is increased, the SVD threshold typically must be decreased correspondingly, resulting in an unsmooth convergence behaviour which cannot be extrapolated. This is a consequence of the more memory efficient representations for the tensors, e.g. matrix product states (MPS).
Furthermore, in cases where calculations are already well converged using the reference method, the optimization provides substantial time savings, often improving efficiency by two orders of magnitude.

The paper is organized as follows: \Sec{System} outlines the general class of systems to which the optimization is directly applicable, and which systems we choose in this paper for illustration. \Sec{PIapproach} summarizes the path-integral formulation for the Trotter decomposition with linked cluster expansion technique, on which the optimization scheme is illustrated. \Sec{OptimisationScheme} introduces the optimization itself on the simplest example of a $2\times2$ reduced density matrix, then generalizes it to an arbitrary number of quantum states.  \Sec{Verification} verifies the optimization scheme by comparing the optimized calculation with calculations via the reference method and the TEMPO algorithm. \Sec{Extrapolation} discusses the extrapolation method which allows us to accurately approximate the exact (infinite neighbors) calculation.

\Sec{Results} is devoted to new physical results. Firstly, we use the optimization scheme to investigate the dephasing rates in a QD-cavity system interacting with acoustic phonons in the regime of very strong exciton-cavity coupling. In this regime, a very strong light–matter interaction is known to dynamically decouple the exciton from its phonon environment \cite{denning_optical_2020}. We show, however, that the resulting decoherence is strongly suppressed but not completely eliminated; in contrast to the interpretation of Ref. \cite{denning_optical_2020}, it remains finite and is instead dominated by virtual phonon-assisted transitions, as demonstrated in this work. Secondly, we show the necessity of the optimization scheme in systems with very long memory times, using for illustration a system of two spatially separated QDs inside a microcavity, in which QD excitons interact to a common phonon bath. In single-QD systems, the memory time is fixed, and one can decrease the time step to achieve a finer time discretization until convergence is achieved. In contrast, in multi-QD systems, the spatial separation determines the memory time.  Since the maximum number of neighbors $L$ is limited by the available RAM, the corresponding time discretization $\Delta t$ may become too coarse at large separations  to accurately capture the quantum dynamics. By applying the optimization and extrapolation, we can reach larger values of  $L$, thereby enabling sufficiently fine time discretization to achieve convergence even in systems with long memory times.

\section{System Hamiltonian}\label{System}
The optimization presented in this paper can be applied to any number of two-level systems (TLSs) that are directly coupled and interact with an environment, either a shared one or independent baths. Additionally, they may interact with an arbitrary number of photonic cavities.
% , the QD-cavity system in \cite{morreauPhononinducedDephasingQuantumdotcavity2019} and the coupled multi qubit-cavity system in \cite{hallControllingDephasingCoupled2024a}.

The general form of the Hamiltonian that is treatable is (in units of $\hbar=1$)
%\begin{equation}\label{H}
%		\begin{split}
%				H =& \; \sum_j \Omega_j d_j^\dagger d_j + \Omega_C a^\dagger a + g (d_1^\dagger d_2 + d_2^\dagger d_1) + \sum_j g_j (d_j^\dagger a + a^\dagger d_j) \\ &+
%				\sum_j d_j^\dagger d_j \sum_{\textbf{q}} \lambda_{\textbf{q},j} (b_{\textbf{q}} + b_{-\textbf{q}}^\dagger)
%				+\sum_{\textbf{q}} \omega_q b_{\textbf{q}}^\dagger b_{\textbf{q}}
%				,
%			\end{split}
%\end{equation}
\begin{equation}\label{H0HIBHph}
	H=H_0 + H_{\mathrm{IB}}\,,
\end{equation}
where $H_0$ describes both the coupling between the TLSs and the TLS-cavity couplings, and is given by
\begin{equation}\label{H0}
	H_0 = \sum_{ij} H_{ij} d_i^\dagger d_j + \sum_k \Omega_{C,k} a_k^\dagger a_k + \sum_{jk} g_{jk} \left( d_j^\dagger a_k + a_k^\dagger d_j \right)\,.
\end{equation}
Here, $H_{ij}$ are the matrix elements of the coupled TLSs, with the diagonal elements ($H_{jj}$) corresponding to the excitation energy of the TLS at site $j$, while the off-diagonal elements ($H_{ij}$, for $i\neq j$) describe the direct coupling between TLSs at sites $i$ and $j$. The fermionic operator $d_j^\dagger$ creates an excitation in the two-level system at site $j$, and a photon in a cavity mode $k$ has energy $\Omega_{C,k}$ and is created by the bosonic operator $a_k^\dagger$. Finally, the TLS at site $j$ is coupled to a cavity mode $k$ with strength $g_{jk}$. In general, there can be multiple baths of 3D bosons, described by
\begin{equation}\label{H_bath}
	H_\text{\rm B}=\sum_l \sum_{\textbf{q}} \omega_{q,l} b_{\textbf{q},l}^\dagger b_{\textbf{q},l}\,,
\end{equation}
where $b_{\textbf{q},l}^\dagger$ creates an excitation in bath $l$ with wave vector $\textbf{q}$.
Within \Eq{H0HIBHph}, $H_{\mathrm{IB}} = H_\mathrm{B}+H_{\mathrm{int}}$ is a generalization of the independent boson model~\cite{mahanManyParticlePhysics2000}, consisting of the free-phonon Hamiltonian $H_\mathrm{B}$ and the TLS-bath interaction
\begin{equation}\label{HIB}
	H_{\textrm{int}} = \sum_j  d_j^\dagger d_j \sum_l\sum_\mathbf{q} \lambda_{\textbf{q},jl} (b_{\textbf{q},l} + b_{-\textbf{q},l}^\dagger)\,,
\end{equation}
where $\lambda_{\mathbf{q},jl}$ describes the interaction strength of the TLS at site $j$ with bath mode $\mathbf{q}$ in bath $l$. In particular, \Eq{HIB} can be used to describe the scenarios where multiple TLSs are coupled to the same bath or coupled to their own independent baths. The diagonal TLS-bath coupling is needed for an exact solvability of the independent-boson Hamiltonian $H_{\mathrm{IB}}$ via linked cluster expansion. However, it has been recently shown that the path-integral based approaches can also be efficiently used for non-diagonal coupling \cite{richterEnhancedTEMPOAlgorithm2022}.

By choosing the number of TLSs and turning on/off specific coupling terms, this general model given by \Eqss{H0HIBHph}{HIB} can be reduced to describe many physical systems, such as energy transport in biological systems~\cite{chinRoleNonequilibriumVibrational2013,reyExploitingStructuredEnvironments2013}, qubits in microwave resonators~\cite{blaisCavityQuantumElectrodynamics2004, zhengPersistentQuantumBeats2013}, quantum dots interacting with a micromechanical resonator \cite{yeoStrainmediatedCouplingQuantum2014}, spin-qubit systems \cite{childressCoherentDynamicsCoupled2006}, and so on.

The specific implementation of the TLSs considered in this paper are semiconductor QDs which are coupled to an environment modeled as a bath of acoustic phonons. Although the general Hamiltonian detailed above describes the range of problems this optimization can treat but is not limited to, we reduce this Hamiltonian to two cases used for illustration. {\em Case 1}: a QD-cavity system coupled to a bath of acoustic phonons detailed in \cite{morreauPhononinducedDephasingQuantumdotcavity2019}; {\em Case 2}: a QD-QD-cavity system coupled to the same phonon bath, detailed in \cite{hallControllingDephasingCoupled2025}. Importantly, in {\em Case 2}, the long memory times are due to the shared bath, where phonons may travel between the QDs. So, the memory time can be as long as the phonon coherence time, and the larger the dot separation, the larger the memory time, which manifests itself as a delay in the bath correlation functions.  For ease of reference, the Hamiltonians for both cases are provided in \App{SysHam}, and details of the exciton-phonon coupling elements and phonon spectral density for these systems are given in \App{App:Coupling}.

\section{Path-integral approach}\label{PIapproach}

For illustration we consider the linear optical response, which we call the linear polarization, as a simple quantum correlator to investigate. However any element of the density matrix, and other quantum correlators may be treated in a similar way, such as the FWM polarization \cite{sirkinaImpactPhononEnvironment2023a} or populations \cite{sirkina_forster_2025}. The linear optical polarization is given by
$P_{jk}(t)={\rm Tr}\{\rho(t) d_j\}$ by definition, where $k$ and $j$ denote, respectively, the excitation channel at $t=0$ and measurement channel at the observation time $t$, and $\rho(t)$ is the full density matrix. As has been derived in Ref.\,\cite{morreauPhononinducedDephasingQuantumdotcavity2019}, the linear polarization can be written as
\begin{equation}\label{LinPol_mt}
	P_{jk}(t) = \langle \bra{j} \hat{U}(t) \ket{k} \rangle_\text{\rm B}\,,
\end{equation}
where $\hat{U}(t) = e^{iH_\text{\rm B}t} e^{-iHt}$ is the evolution operator and $\langle...\rangle_\text{\rm B}$ denotes the expectation value over all phonon degrees of freedom in thermal equilibrium.
Common to path-integral based approaches, the time interval $[0,t]$, where $t$ is the observation time, is split into $N$ equal steps of duration $\Delta t = t/N = t_n - t_{n-1}$, where $t_n = n\Delta t$ represents the time at the $n$-th step. Trotter's theorem is then used to separate the time evolution of the two non-commuting components of the Hamiltonian, ${H}_0$ and $H_\text{\rm IB}$. In fact, applying Trotter's decomposition theorem, the time evolution operator $\hat{U}(t)$ can be written as
\begin{equation}\label{TrottersU_mt}
	\hat{U}(t) = \lim_{\Delta t \to 0} e^{iH_\text{\rm B}t}(e^{-iH_\text{\rm IB}\Delta t} e^{-iH_0\Delta t})^N\,.
\end{equation}
The final exponential in \Eq{TrottersU_mt} describes the dynamics of the system in the absence of phonons and is written as an operator $\hat{M}$:
\begin{equation}\label{Moperator}
	\hat{M}(t_n-t_{n-1}) = \hat{M}(\Delta t) = e^{-iH_0\Delta t}.
\end{equation}

\begin{figure}[t]
	\begin{tabular}{c} % Stack figures vertically
		% First row: G tensor
		\parbox[b]{0.05\textwidth}{ \raisebox{2ex}{\hspace*{-2em}$\mathcal{G}:$}}%
		\parbox[b]{0.3\textwidth}{\centering \includegraphics[width=0.3\textwidth]{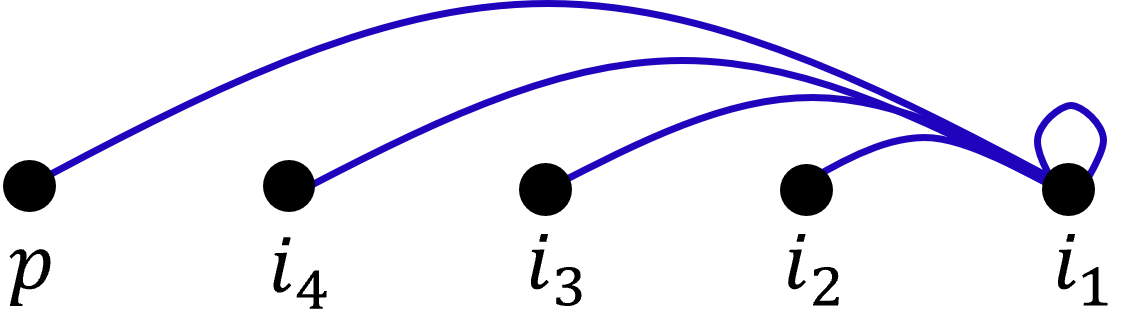}}%
		\\
		% Second row: F tensor
		\parbox[b]{0.05\textwidth}{\raisebox{2ex}{\hspace*{-2em}$F:$}}%
		\parbox[b]{0.3\textwidth}{\centering \includegraphics[width=0.24\textwidth]{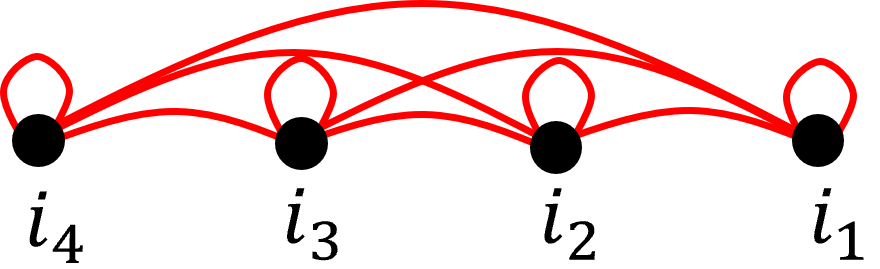}}%
	\end{tabular}
	
	% Single caption for both images
	\caption{Diagrams showing the path segments contained within $\mathcal{G}$ (blue) and $F$ (red) for $L=4$. The propagator $\mathcal{G}$ contains path segments connecting the given time (indexed by $i_1$) to itself and to all other considered nearest time intervals (neighbors, indexed by $i_n$). Each link is weighted appropriately, forming a memory kernel. The full influence functional $F$ represents the information about the state of the system and contains all considered path segments and all possible links between them.}
	\label{tensor_comparison}
\end{figure}

However, the necessity of the path-integral approach arises due to the interaction of the system with the bath within $H_\mathrm{IB}$ which can be handled in various ways but always results in a tensor multiplication scheme of some form, which the present optimization scheme can be used for. In particular, we choose to apply linked cluster expansion \cite{mahanManyParticlePhysics2000}.  This leads to the tensor multiplication scheme used in ~\cite{morreauPhononinducedDephasingQuantumdotcavity2019,sirkinaImpactPhononEnvironment2023a,hallControllingDephasingCoupled2025, sirkina_forster_2025}:
\begin{equation}
	F_{p i_L \ldots i_2}^{(s+1)} = \sum_{i_1} \mathcal{G}_{p i_L\ldots i_1} F_{i_{L}\ldots i_1}^{(s)}\,,
	\label{Fn}
\end{equation}
where $\mathcal{G}$ is known as the propagator and $F^{(s)}$ as the full influence functional, and \Fig{tensor_comparison} shows the path segments contained in the tensors. Key details of the approach are outlined in \App{App:Trotter-Linked}. The role of $\mathcal{G}$ is to propagate the system forward in time, taking the tensor $F^{(s)}$ to $F^{(s+1)}$, where $F$ contains all correlations within the memory time. In particular, $F$ captures how past states influence the present dynamics, incorporating non-Markovian effects.
The number of correlations, or time steps, included in the $F$ tensor is also referred to as the number of neighbors, $L$, with \Fig{tensor_comparison} depicting $4$ neighbors. Each index $i_n$ of the tensor can have $J$ possible values, corresponding to $J$ different quantum states within the reduced density matrix. For example, in {\em Case 2}  of a QD-QD-cavity system,  $J=3$, where we define $i_n= 1$ or $i_n=2$ to correspond to the excitonic channels in QD $1$ or $2$, respectively, and $i_n=0$ represents the cavity channel. At any time step $n$ within the memory kernel, the excitation can transfer between these components, meaning that the index $i_n$ within the tensors must account for the fact that the system oscillates between QD $1$, QD $2$, and the cavity.

In \Fig{tensor_comparison}, $\mathcal{G}$ shows a specific case of two-time correlations, employed in all of the path-integral techniques mentioned so far. This is a consequence of the assumption that the system-bath coupling is bilinear in system operators (and linear in the bath operators), in which case all higher-order correlation functions can be expressed in terms of the two-time correlations. In the present case of the exactly solvable IB model describing the system-bath interaction, the cumulant expansion gives the exact result in its lowest (second) order, so $\mathcal{G}$ is given by
\begin{equation}
	\label{GtensorLN_mt}
	\mathcal{G}_{p \dots i_1} = M_{i_2 i_1}  e^{\mathcal{K}_{i_1 i_1}(0) + 2\mathcal{K}_{i_2 i_1}(1) + 2\mathcal{K}_{i_3 i_1} (2)+ \dots + 2\mathcal{K}_{p i_1}(L)  }\,,
\end{equation}
where $\mathcal{K}$ denotes a cumulant arising from the application of linked cluster expansion \cite{mahanManyParticlePhysics2000}. The cumulants contain two indices, $i_n$ and $i_m$, describing correlations between two time points $t_n$ and $t_m$, and take into account the fact that not all components of the system couple to phonons.  More details on the linked cluster expansion and explicit form of the cumulants can be found in \App{App:Trotter-Linked}, with full details provided in Refs.\,\cite{morreauPhononinducedDephasingQuantumdotcavity2019,hallControllingDephasingCoupled2025}.

The initial full influence functional at the first time step is simply given by $F_{i_L\ldots i_1}^{(1)}= M_{i_1 k}$, where $k$ is the excitation channel and $M_{ij}$ are the matrix elements of the operator $\hat{M}$ defined in \Eq{Moperator}.
The final full influence functional $F_{i_L\ldots i_1}^{(N)}$ at the observation time $t=N\Delta t$ is found by applying the propagator $N-1$ times, according to \Eq{Fn}, and in general determines the full reduced density matrix of the system at time $t$. The off-diagonal elements of this density matrix yield the
linear optical polarization, which is given by
\begin{equation}\label{PjkLN}
	P_{jk}(t) = e^{\mathcal{K}_{jj}(0)} F_{0\dots 0 j}^{(N)}\,.	
\end{equation}
Here, the subscripts $0\dots 0$ represent the indices placed into the cavity channel, which is not coupled to phonons, but more generally, this can be any channel decoupled from the bath, such as the absolute ground state of the system. Note also that $k$ is incorporated implicitly within the construction of the initial tensor $F_{i_L \ldots i_1}^{(1)}$ in \Eq{Fnm}, see below.

As the number of time steps within the memory kernel increases (increasing neighbors), the tensors in \Eq{Fn} are growing exponentially in size. The exponential growth limits the number of correlations that can be considered, due to computational limitations. As a consequence there are problems in achieving convergence for the following scenarios: (i) When the system has long phonon memory times; (ii) when there is a large number $J$ of quantum states or channels; (iii) in strong coupling regimes where the quantum dynamics of the system exhibits rapid oscillations.

\section{Optimization scheme}\label{OptimisationScheme}
For clarity of presentation, we first introduce in \Sec{Sec:J2} the optimization scheme for {\em Case 1}, which is the linear polarization in a QD-cavity system, with the QD exciton (the excited state of the TLS) coupled to a bath of acoustic phonons. This is the simplest possible non-trivial system treated by the Trotter decomposition method with cumulant expansion \cite{morreauPhononinducedDephasingQuantumdotcavity2019}, which requires two basis states ($J=2$) such that $i_n=0$  ($i_n=1$) indicates the system is in the cavity (exciton) state at time step $n$.
While we present below the optimization for an arbitrary number of neighbors $L$ within the so-called $L$-neighbor approach, all the figures in \Sec{Sec:J2} use only $L=4$ neighbors as a simpler example to ease understanding. A generalization of the optimization scheme for an arbitrary number $J$ of basis states or elements of the density matrix is discussed in \Sec{Sec:generalJ}.

\subsection{Optimization for the smallest reduced density matrix, $J=2$}
\label{Sec:J2}

The core principle of the optimization scheme is to map the tensors in \Eq{Fn}, in particular the full influence functional $F$,
into two matrices and then to SVD these matrices at each time step, in order to minimize their size by filtering out contributions below a desired threshold value.

Let us consider the first time step, after excitation in channel $k$ at $t=0$. In this case, the full influence functional tensor can be fully populated by $M_{i_1 k}$, given by \Eq{Moperator} and then mapped onto a matrix $\mathcal{F}_{nm}$:
\begin{equation}
	\label{Fnm}
	F_{i_L\ldots i_1}^{(1)} = M_{i_1 k} = \mathcal{F}_{nm}^{(1)}\,.
\end{equation}
The full influence functional is populated by only two values of $M_{i_1 k}$: $M_{0k}$ or $M_{1k}$. In fact, following the excitation at $t=0$ in channel $k$, one further time step introduces index $i_1$, which has two possible paths of evolution, $i_1=0$ and $i_1=1$. However, the tensor $F_{i_L\ldots i_1}^{(1)}$  depends on $L$ indices $i_L\ldots i_1$,  and consequently the matrix $\mathcal{F}_{nm}^{(1)}$ also contains the information about all of them in some form, as detailed below. This is already different from other methods, such as Ref.\,\cite{strathearnEfficientNonMarkovianQuantum2018}, in which there is always a growth phase in the early times, where each new time step introduces a new index to the tensor, up until the full memory kernel has been formed, and beyond this point the tensor has a fixed maximum number of indices. We begin instead with the full-rank tensor as a placeholder for subsequent propagation.

\begin{figure}[t]
	\centering
	\includegraphics[scale=0.6]{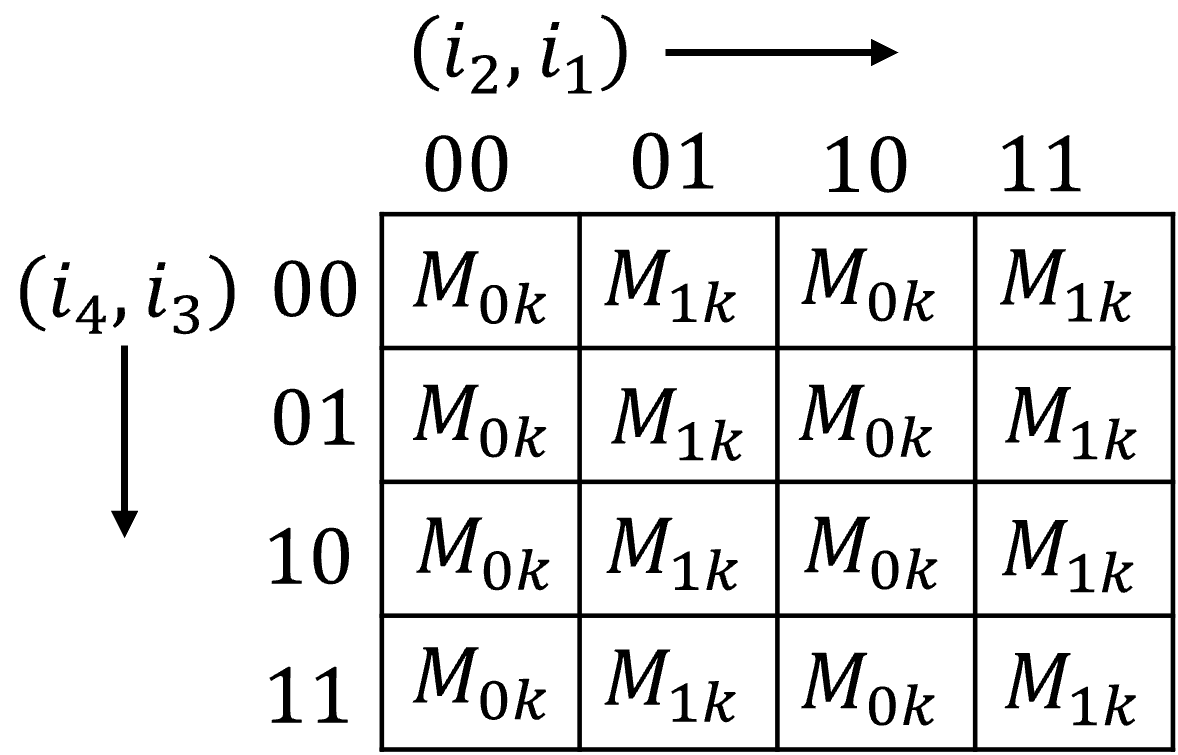}% Here is how to import EPS art
	\caption{\label{Fnm_initial} The tensor $F_{i_4 i_3 i_2 i_1}^{(1)}$ mapped onto a matrix $\mathcal{F}_{nm}^{(1)}$, with the column index corresponding to the possible values of $(i_2, i_1)$ and the row index to $(i_4,i_3)$. }
\end{figure}
\begin{figure}[t]
	\centering
	\includegraphics[scale=0.6]{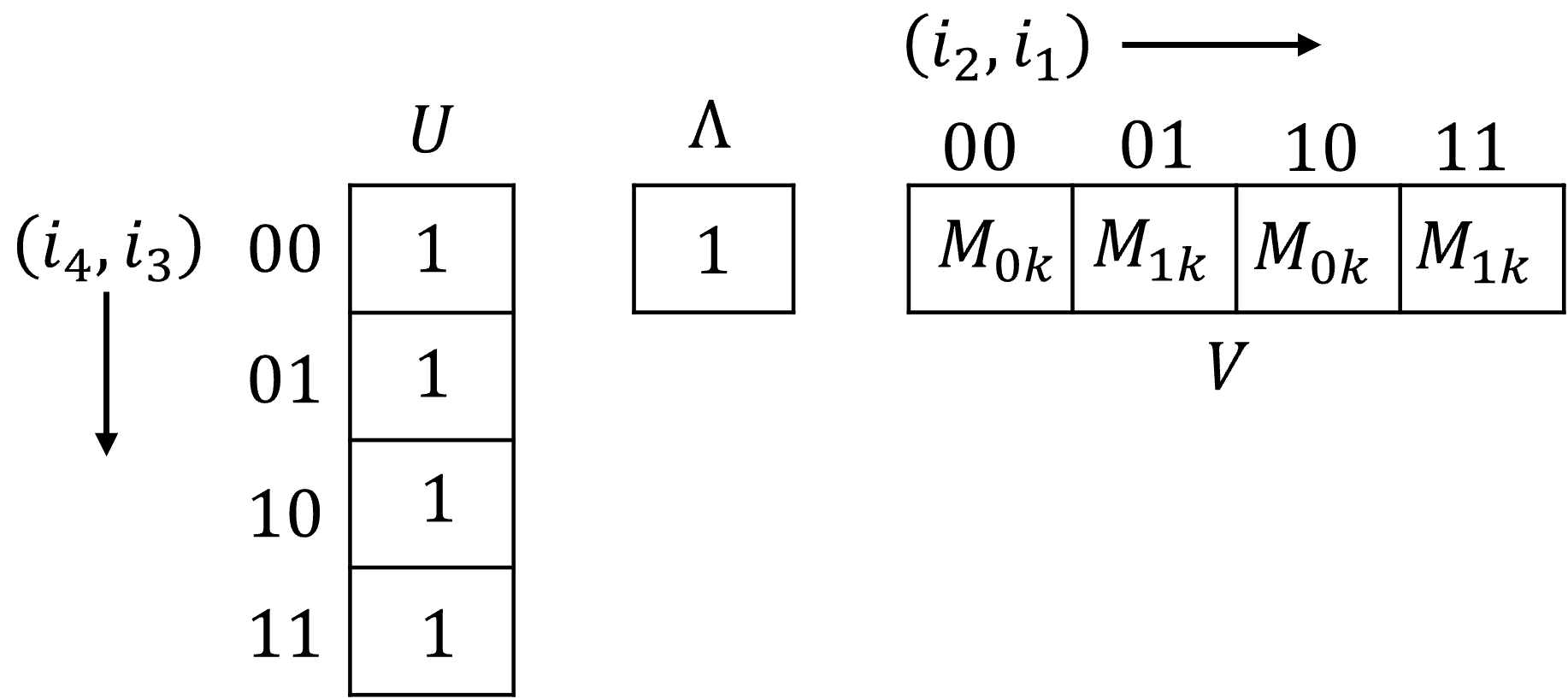}% Here is how to import EPS art
	\caption{\label{F_initial_SVD} The matrix in \Fig{Fnm_initial}, $\mathcal{F}_{nm}^{(1)}$, expressed in SVD form.}
\end{figure}

\Fig{Fnm_initial} depicts the $L=4$ influence functional $F$ remapped to a matrix $\mathcal{F}_{nm}$.
In $\mathcal{F}_{nm}$, the columns take into consideration the possible values of the indices $i_{\frac{L}{2}}\ldots i_1$, and the rows add the remaining indices $i_L \ldots i_{\frac{L}{2}+1}$. Note that in general $L$ does not need to be even, but below we assume that it is even for clarity of presentation. Since the linear polarization in the QD-cavity system only has the possible index options $i_n=0$ or $1$, for $L/2$ indices associated with each of the two matrices, there are $2^{L/2}$ index configurations to take into account. We generate all possible index configurations by enumerating all tuples \( (i_\frac{L}{2}, \dots, i_1) \), where each index \( i_n \) independently takes values $0$ or $1$.
Each configuration represents a possible path of the system as it evolves over the time steps. For $L=4$, the configurations for each pair of indices $i_2i_1$ and $i_4i_3$ are simply $\{$00,01,10,11$\}$. Thus, the dimensions of the mapped matrix $\mathcal{F}_{nm}$ are $n_\textrm{max}$ and $m_\textrm{max}$, where $n_\textrm{max}=m_\textrm{max}=2^{L/2}$ in general.
% for even $L$ (or $n_\textrm{max}/2=m_\textrm{max}=2^{L/2}$ for odd $L$).
However, $\mathcal{F}_{nm}$ contains the same number of elements as the original tensor $F$, and therefore presently provides no memory storage reduction.
To remedy this, in order to avoid constructing large tensors in the first instance, $\mathcal{F}^{(1)}_{nm}$ can be analytically expressed in SVD form. In general, this is always possible to do, and is given by
\begin{equation}
	\label{Fnm_svdinitial}
	\mathcal{F}_{nm}^{(1)} =  U_{n0} \, \Lambda_0 \, V_{0m}\,,
\end{equation}
with $U_{n0}=1$, $\Lambda_0 =1$ and $V_{0m}=M_{i_1 k}$. Here $\Lambda$ is the $1\times 1$ identity matrix, but it later grows in size and is used for filtering out unimportant contributions after performing a SVD. As an example, \Fig{F_initial_SVD} shows the matrix $\mathcal{F}_{nm}^{(1)}$ in such an SVD form for $L=4$. This reduces the total number of elements from $2^{L}$ to $2^{L/2 +1}$, having a significant impact at larger $L$, approximately doubling the amount of neighbors achievable. As the matrix has now been represented in SVD form, we define a column vector $U$ which takes into account the indices $i_L\ldots i_{\frac{L}{2}+1}$ and the row vector $V$, taking into account indices $i_\frac{L}{2}\ldots i_1$.

For further time steps, the exciton-phonon coupling has to be considered, which is taken into account via the propagator $\mathcal{G}$ in \Eq{Fn}. $\mathcal{G}$ is successively applied to propagate the system forward in time, summing over the first index, $i_1$, as seen in \Eq{Fn}.
Although the propagator \Eq{GtensorLN_mt} is in the form of a tensor, the two-time correlations can be expressed as $2\times2$ matrices in the following way
\begin{align}
	\label{Qii1}
	Q_{p \, i_1}^{(L)} &= \exp\left\{2\mathcal{K}_{p\, i_1}(L)\right\}  \,, \nonumber \\
	Q_{i_r \, i_1}^{(r-1)} &= \exp\left\{2\mathcal{K}_{i_{r}\, i_1}(r-1)\right\} \quad \text{for }
2 < r \leqslant L\,, \nonumber \\
	Q_{i_2\, i_1}^{(1)} &= M_{i_2 i_1} \exp\left\{ \mathcal{K}_{i_1 i_1}(0) + 2\mathcal{K}_{i_2 i_1}(1) \right\}
	\,,
\end{align}
where the superscript denotes the time step difference between any index (not including $i_1$) in the memory kernel and $i_1$, i.e. the time step difference between the indices within each pair correlation. For example, considering the indices $i_3$ and $i_1$, the correlations are between time points $t_3$ and $t_1$ in the memory kernel, thus the $Q$ matrix describing these correlations is given by $Q_{i_3i_1}^{(2)}$.
The recursive relation \Eq{Fn}, can then be re-expressed as
\begin{equation}
	F_{p \, i_L\ldots i_2}^{(s+1)} = \sum_{i_1} Q_{p\, i_1}^{(L)} Q_{i_L i_1}^{(L-1)} \ldots Q_{i_3 i_1}^{(2)} Q_{i_2 i_1}^{(1)} \mathcal{F}_{nm}^{(s)}\,.
	\label{Fn_Qs}
\end{equation}

The product of the $Q$ matrices must be applied to the appropriate elements of $\mathcal{F}_{nm}^{(s)}$, with a summation over index $i_1$. At any time step $s$, $\mathcal{F}_{nm}^{(s)}$ has the SVD form
\begin{equation}
	\label{Fnm_svds}
	\mathcal{F}_{nm}^{(s)} =  \sum_k U_{nk} \, \Lambda_k \, V_{km}
\end{equation}
by construction, as it is clear from the following, which includes as a special case the initial, $s=1$ matrix given by \Eq{Fnm_svdinitial}. Since the matrix $U$ ($V$) takes into account the indices $i_L \ldots i_{\frac{L}{2}+1}$ ($i_\frac{L}{2}\ldots i_1$), only the $Q$ matrices which introduce pair correlations between those indices and the index $i_1$ should be applied to that matrix. Additionally, matrix $Q_{p i_1}^{(L)}$ bringing in a new index $p$ is applied to $U$
We then obtain two new matrices,
\begin{align}
	\tilde{U}_{nk}^{(i_1)}
	&= Q_{i_L i_1}^{(L-1)} \cdots Q_{i_{\frac{L}{2}+1} i_1}^{(\frac{L}{2})} \, U_{nk} \,, \\
	\tilde{V}_{k\tilde{m}}^{(i_1, p)}
	&= Q_{p i_1}^{(L)} Q_{i_{\frac{L}{2}} i_1}^{(\frac{L}{2}-1)} \cdots Q_{i_2 i_1}^{(1)} \, V_{k m}\,,
	\label{UVtilde}
\end{align}
where the difference between $m$ and $\tilde{m}$ is that $m$ includes index $i_1$ but $\tilde{m}$ does not, as it is now used as a superscript in $\tilde{V}$. Note that as the left matrix $U$ does not contain any information about $i_1$, this index appears as an addition superscript in $\tilde{U}$. Similarly, the new index $p$ in the propagator $\mathcal{G}$ appears as another superscript in $\tilde{V}$, in addition to $i_1$ which was intentionally separated from all other indices as there is a summation over $i_1$. Now, having applied all the propagator matrices, the tensor $F_{p \, i_L\ldots i_2}^{(s+1)}$ in \Eq{Fn_Qs} remapped to a matrix given by
\begin{equation}
	\label{Fnm_psplit}
	{\mathcal{F}}_{nm}^{(s+1)} = \sum_{i_1} \sum_k \tilde{U}_{nk}^{(i_1)} \Lambda_k \tilde{V}_{k\tilde{m}}^{(i_1,p)}
\end{equation}
and expressing propagation of the system forward in time, where the difference between indices $m$ and $\tilde{m}$ is that $m$ now includes the new index $p$ but $\tilde{m}$ does not.

\begin{figure}[t]
	\centering
	\includegraphics[scale=0.5]{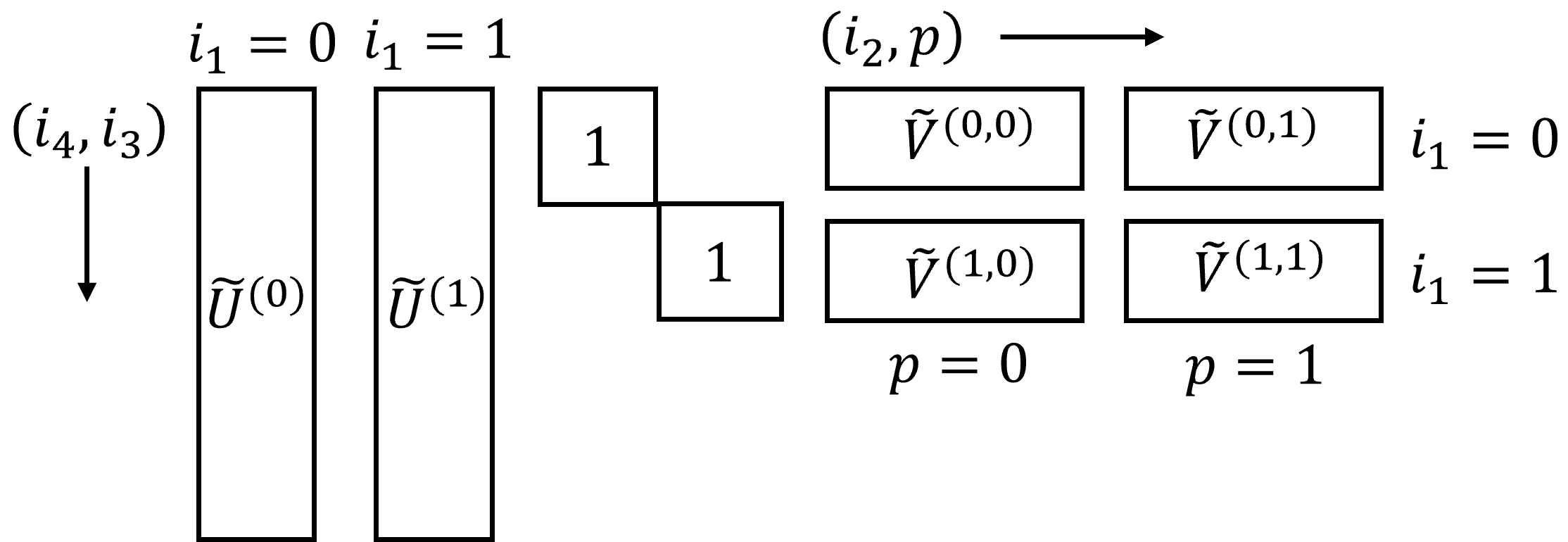}% Here is how to import EPS art
	\caption{\label{F_s+1} The tensor $F_{p \, i_L\ldots i_2}^{(2)}$ mapped onto a matrix ${\mathcal{F}}_{nm}^{(2)}$. The vectors from the initial time step, $U_{n0}$ and $V_{0m}$, split through applying the $Q$ matrices and the elements with $i_1=0$ and $i_1=1$ separated. The blocks of $\tilde{V}$ are further split due to the extra index $p$, with $p=0$ and $p=1$ separated. }
\end{figure}

With reference to the $L=4$ special case illustrated in \Fig{F_initial_SVD} for $s=1$, the propagator matrices $Q_{i_2\,i_1}^{(1)}$ and $Q_{p\,i_1}^{(4)}$ are applied to the right vector $V$, and matrices $Q_{i_3\,i_1}^{(2)}$ and $Q_{i_4\,i_1}^{(3)}$ to the left vector $U$. As there is no information about index $p$ contained within either left or right vectors, both possibilities of $p=0$ and $1$ must be taken into account. \Fig{F_s+1} shows the $s=2$ newly propagated matrix ${\mathcal{F}}_{nm}^{(2)}$ for $L=4$, where $\tilde{U}$ and $\tilde{V}$ have been split into blocks due to the summation over $i_1$ in \Eq{Fnm_psplit}, and according to \Eq{UVtilde}, the $Q$ matrices have been applied, further splitting $\tilde{V}$ into four blocks due to the introduction of a new index $p$ in the propagator.

\begin{figure}[t]%!h]
	\centering
	\includegraphics[scale=0.6]{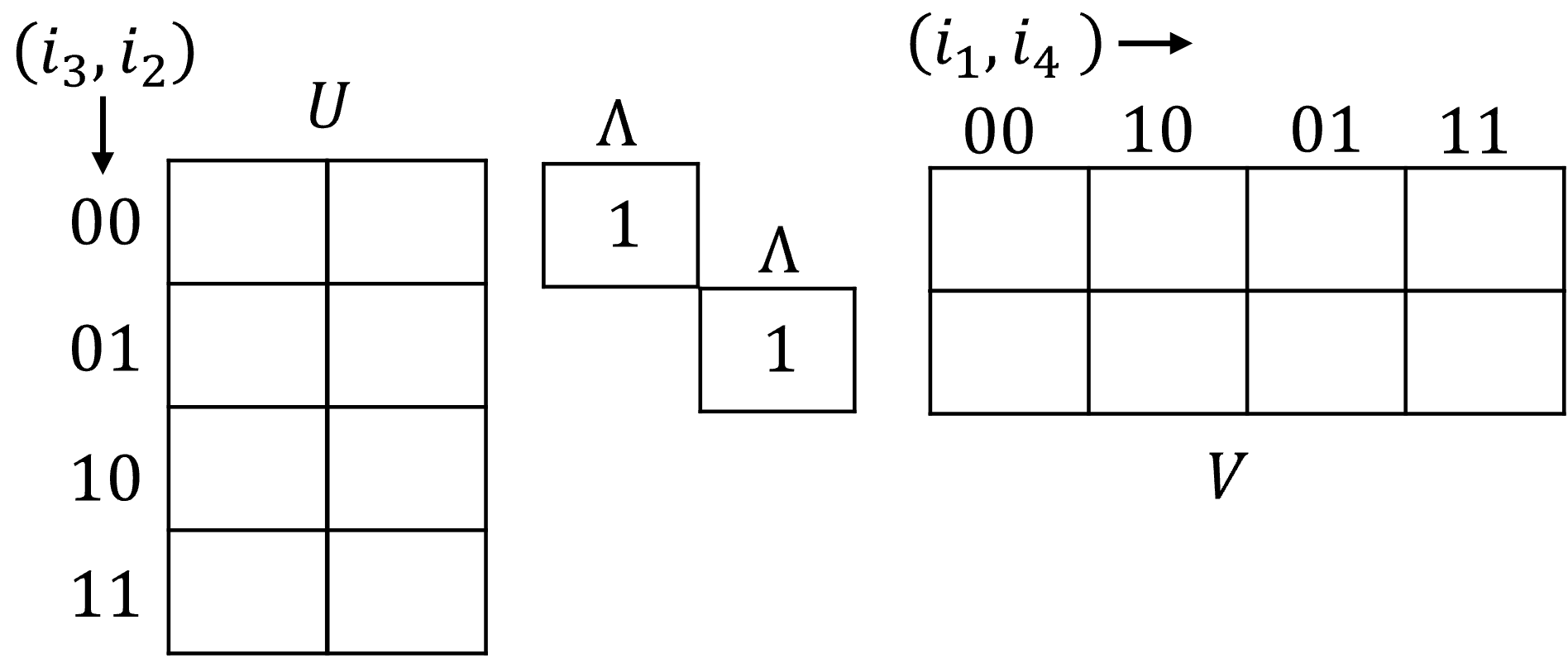}% Here is how to import EPS art
	\caption{\label{U_V_secondstep} The tensor $F_{i_L\ldots i_1}^{(2)}$ mapped onto a matrix $\mathcal{F}_{nm}^{(2)}$, with matrices $U$ and $V$ redefined and the indices relabeled. This is the starting point of the next time step, in which the subsequent application of the $Q$ matrices and splitting due to $i_1=0,1$ and $p=0,1$ will be applied again to propagate the system. }
\end{figure}

In the recursion relation \Eq{Fn} [or \Eq{Fn_Qs}], $F_{p \, i_L\ldots i_2}^{(s+1)}$ must become the new $F_{i_L \ldots i_1}^{(s)}$ for the next time step, and we must therefore relabel the indices. As an example, for $L=4$, relabeled indices $i_4i_3 \rightarrow i_3i_2$ becomes associated with $\tilde{U}$ and $i_2 p \rightarrow i_1 i_4$ with $\tilde{V}$. In addition to this change of indices, we rename the matrices  $\tilde{U}$ and $\tilde{V}$ back to  $U$ and $V$. In other words, after each time step, the blocks $\tilde{U}$ and $\tilde{V}$ form the new $U$ and $V$, as illustrated in \Fig{U_V_secondstep}, having the same size in the dimensions associated with the new indices as the old matrices but generally different size in the other dimension, given by the indices of summation $i_1$ and $k$ in \Eq{Fnm_psplit}.

Applying \Eq{Fn_Qs} further, in each successive time step, the indices will cycle, but once the index $i_1$ becomes associated with $U$, which occurs after $L/2$ steps, the original index ordering has been restored but on opposite matrices. Specifically, $i_{L/2} \ldots i_1$ is now instead associated with $U$ and $i_{L} \ldots i_{\frac{L}{2}+1}$ with $V$. Once this occurs, it is computationally simpler to transpose all the matrices and redefine, such that $V^T\to U$ and $U^T\to V$, allowing us to repeat the procedure outlined in this section above.

\begin{figure}[t]
	\centering
	\includegraphics[scale=0.6]{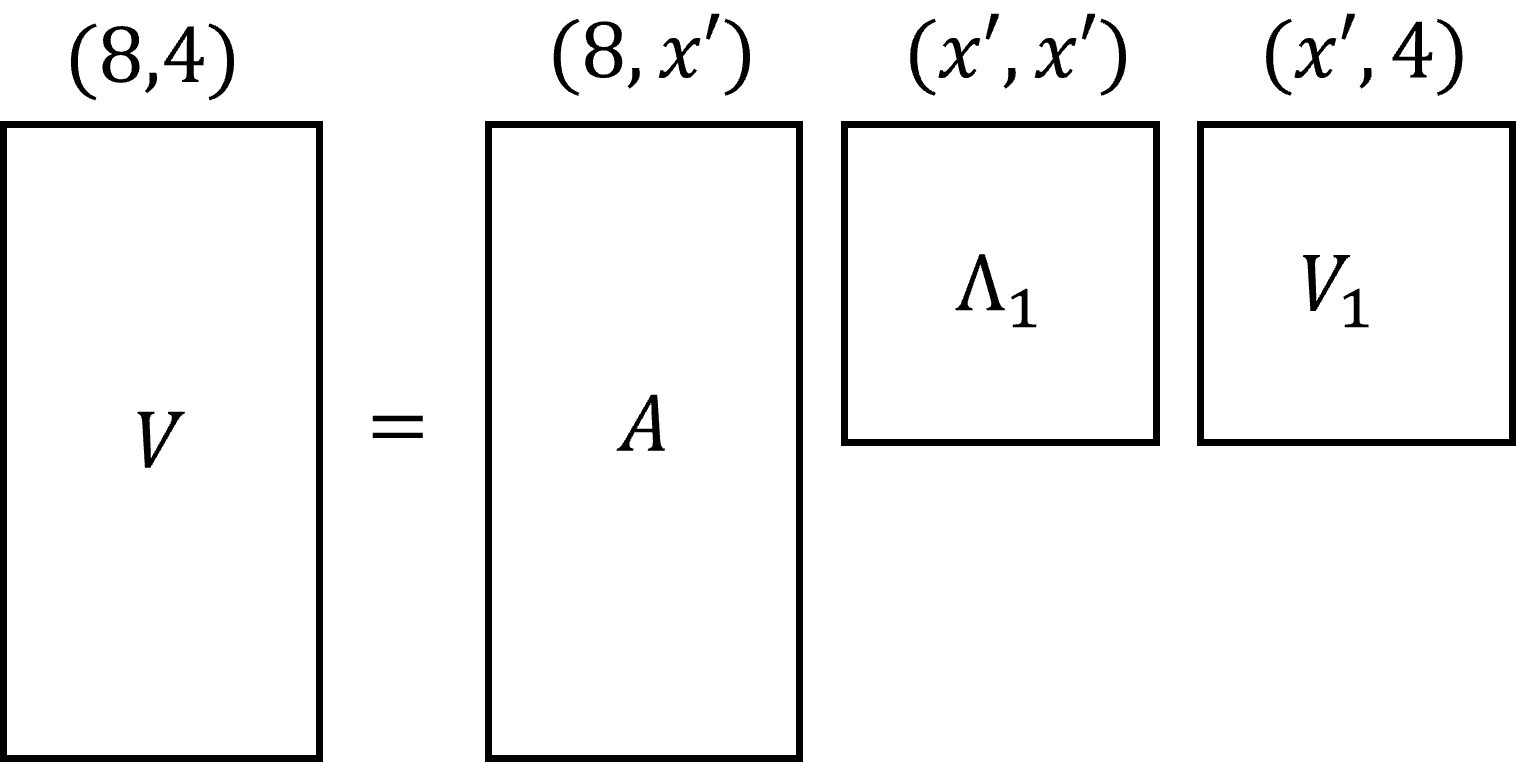}% Here is how to import EPS art
	\caption{\label{SVD_on_V} SVD applied on $V$ after the third time step ($s=4$), where there are now $8$ rows. The number of columns (4 for $L=4$) is fixed due to the fixed number of configurations $2^{L/2}$ for a given number of neighbors $L$. By applying SVD with a threshold value, the truncated dimension becomes $x'$. The truncated matrix $V_1$ replaces $V$ and is used for further propagation. }
\end{figure}

It is clear that through the splitting due to $i_1=0$ and $1$ after each time step, the size of $U$ matrix doubles in size relative to the previous step, and through the splitting due to $p=0$ and $1$, $V$ also doubles in size. %Thus, every subsequent $\mathcal{F}$ doubles in size,
However this can be optimized, reducing the sizes, by SVDing the matrices.
For instance, we can SVD $V$, as shown in \Fig{SVD_on_V}, where $\Lambda_1$ which contains the significant values has a smaller size due to the threshold condition that if $\lambda_k < \epsilon \lambda_{\text{max}}$ then $\lambda_k = 0 $, where $\lambda_\text{max}$ is the largest value contained in $\Lambda_1$.  Clearly, this truncation of the SVD matrices does not affect their dimensions associated with the indices $i_L\ldots i_1$.  The threshold value $\epsilon$ may be chosen based on the desired accuracy of the calculation, and in this paper we use $\epsilon=  10^{-8}$.
As seen in \Fig{SVD_on_V}, for illustration considering $V$, which is initially a $1\times4$ matrix at the first time step, but has increased to an $8\times 4$ matrix after three iterations (but would be an $8\times 2^{L/2}$ in general, for $s=4$). Then, if after applying the threshold only $x'$ significant values remain, $A$ is an $8\times x'$ matrix, $\Lambda_1$ is an $x'\times x'$ matrix, and $V_1$ is an  $x'\times 4$ matrix.
 \begin{figure}[t]
 	\centering
 	\includegraphics[scale=0.6]{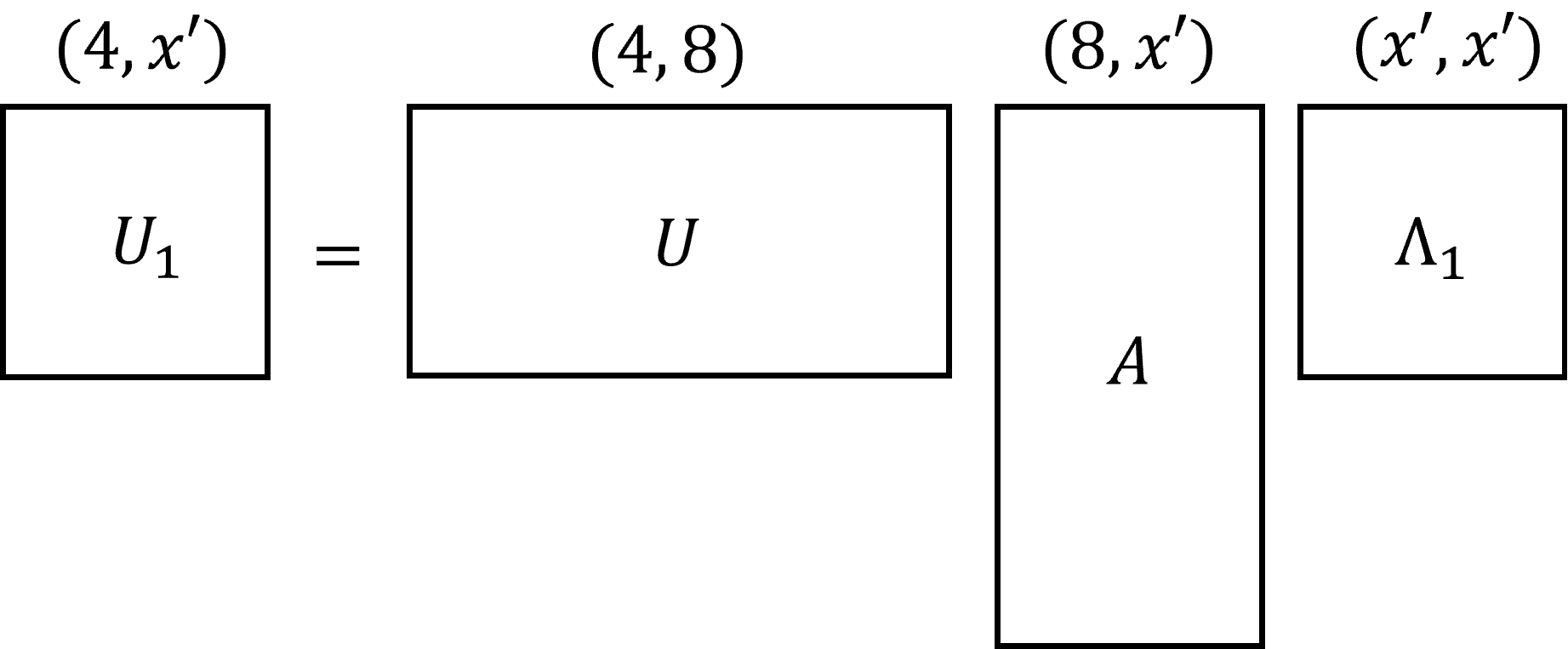}% Here is how to import EPS art
 	\caption{\label{SVD_on_U} The matrices $A$ and $\Lambda_1$ from SVD of $V$, multiplied by $U$ to form $U_1$, which automatically truncates the dimensions without performing any further SVDs.
 The reduced matrix $U_1$ replaces $U$ and is used for further propagation.
 }
 \end{figure}
Initially this appears to increase the number of required elements, however $A$ and $\Lambda_1$ can be multiplied into $U$, which is a $4\times8$ matrix, and through matrix multiplication, the size of $U$ is truncated to a matrix $U_1$ of size $4\times x'$. Note that once the filtering of unimportant contributions from $\Lambda_1$ has occurred, $\Lambda_1$ can be absorbed into either $U_1$ or $V_1$. Thus, the matrices of reduced dimension, $V_1$ and $U_1$ in Figs.\,\ref{SVD_on_V} and \ref{SVD_on_U}, respectively, are used as the new $V$ and $U$ for the following time step. One can introduce an identity matrix of dimension $x' \times x'$ between the new $U$ and $V$ to maintain the SVD form in \Eq{Fnm_svds}. If the truncation is insufficient after one SVD, one can apply SVD again, this time to the new matrix $U$, and then multiply the SVD blocks from $U$ back into $V$. Effectively, there is an SVD sweeping back and forth to truncate the sizes. These truncated matrices are then used as the starting $U$ and $V$ for the following time step, with the SVD procedure applied every time step to prevent exponential growth of matrices.

\subsection{Generalization to J-dimensional density matrix vectors}
\label{Sec:generalJ}

The procedure described in \Sec{Sec:J2} for $J=2$ is general and computationally straightforward. It can therefore be applied to the reduced density matrix of any size, or to an arbitrary number of quantum states $J$ describing the system. To summarize it, we start with a tensor which has a number of indices equal to the number of time steps, or neighbors, $L$, within the memory kernel. We then assign about a half of the indices to a matrix $U$, and the rest to a matrix $V$, with each configuration of the indices corresponding to a specific path of quantum evolution. Due to the exponential form of the propagator, it can be decomposed into matrices $Q$, defined in \Eq{Qii1}, which describe the correlations between two time points. The $Q$-matrix elements can be multiplied into $U$ and $V$, propagating the system. Then, via matrix multiplication of the desired rows and columns, any physical observable can be found. For example, the linear polarization in {\em Case 1} or {\em 2} can be found via \Eq{PjkLN}. The exponential scaling of the matrix sizes (doubling at each time step for $J=2$) is then handled by SVDing the matrices to keep only contributions above a chosen threshold value.

\begin{figure}[t]%!h]
	\centering
	\includegraphics[scale=0.6]{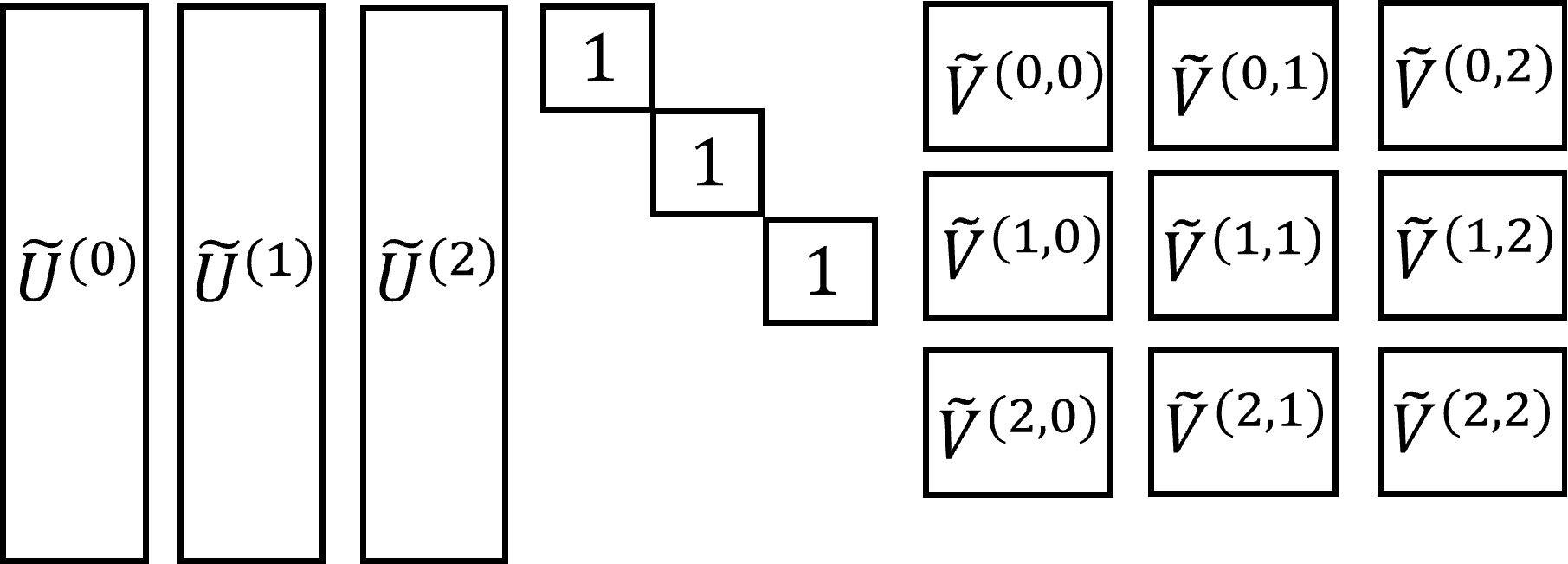}
	\caption{\label{QD-QD-cav-F}
As \Fig{F_s+1} but for $J=3$, corresponding to {\em Case 2}, the QD-QD-cavity system coupled to the phonon bath.
%The tensor $F_{p \, i_L\ldots i_2}^{(s=2)}$ mapped onto a matrix ${\mathcal{F}}_{nm}^{(2)}$ for $J=3$, as in {\em Case 2} for the QD-QD-cavity system. The matrix $U$ has been duplicated three times to account for all possible values of $i_1=0, 1, 2$, since no knowledge of $i_1$ is contained within $U$. $V$ has been split into three blocks corresponding to the values of $i_1$, as the value of $i_1$ is known for each element in $V$. However, each block of the split $V$ is then duplicated three times due to the propagation index $p=0, 1, 2$.
}
\end{figure}

Here we illustrate the more general case for a system and correlator which can be fully described by $J$ system states, or elements of a reduced density matrix. At time step $s$, the tensor $F_{i_L \ldots i_1}^{(s)}$ (where $i_n = 0,1,\ldots J-1$) is represented by a standard matrix $\mathcal{F}_{nm}^{(s)}$, which is split into two matrices $U$, $\Lambda$ and $V$ as before, see \Eq{Fnm_svds}. Here $\Lambda$ is simply the identity matrix for steps $s=1$ and $s=2$, but is responsible for truncation after applying SVD and filtering out unimportant contributions for any further step ($s>2$). The indices are assigned to the matrices $U$ and $V$, and after applying the $Q$ matrices, blocks of $\tilde{U}$ and $\tilde{V}$ are formed. One main difference to the previous  section dealing with $J=2$ is that the indices can be any of the $J$ possible system states, rather than simply $0$ or $1$. Thus, the dimensions of the mapped matrix, $n_\textrm{max}$ and $m_\textrm{max}$, are dependent on $J$ and the number of neighbors $L$: $n_\textrm{max}=m_\textrm{max} = J^{L/2}$ (for even $L$). In other words, there are more configurations of the indices when there are more possible system states.
%due to the extra possible states of $i_1$, the matrix $V_{km}$ is split into $J$ matrices, $V_{km}\rightarrow V_{km'}^{(i_1)}$, each corresponding to a specific $i_1$ value. $V_{km'}$ is a $J$ times smaller matrix than $V_{km}$ since the $i_1$ values are known, as this index is associated with $V$. The propagation index $p$ also has $J$ possible values, and each $V_{km'}^{(i_1)}$ is duplicated to account for all possible $p$ values. Similarly, $U_{nk}$ is duplicated $J$ times, with each corresponding to a possible $i_1$ value, as there is no information about index $i_1$ contained within $U$.

The other main difference is in the number of blocks of $\tilde{U}$ and $\tilde{V}$ after applying the Q matrices, defined in \Eq{Qii1}. \Fig{QD-QD-cav-F} shows the specific case of $J=3$, which may represent {\em Case 2}, the linear polarization in a QD-QD-cavity system, in which case the indices take the values $i_n = 0,\,1$, and  2. It illustrates the effect of the $Q$ matrices, which are applied to the matrix elements of $V$ and $U$, as described in \Sec{Sec:J2}, generally resulting in $J$ blocks of $\tilde{U}$ and $J^2$ blocks of $\tilde{V}$, each of them being $J$ times shorter than $V$.

\section{Verification}\label{Verification}
\begin{figure}[t]
	\centering
	\includegraphics[width=0.5\textwidth]{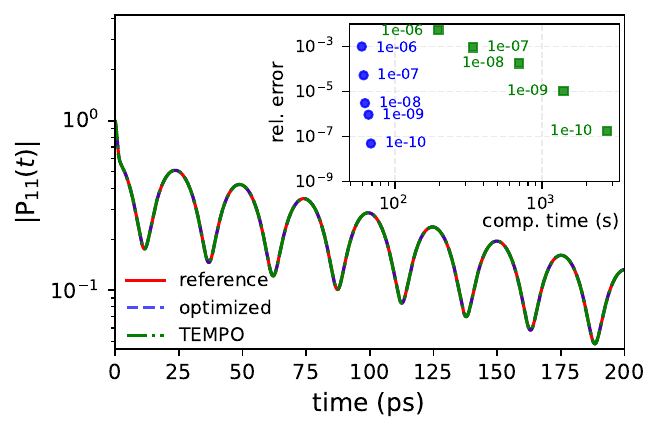}
	\caption{\label{QDCAVITY_Comparison}
		Linear optical polarization $|\mathrm{P}_{11}(t)|$ of an isotropic QD with the Gaussian confinement length $l=3.3\,$nm, coupled to a microcavity with coupling strength $g=100\,\mu$eV at zero detuning, with excitation and measurement in the QD. The calculation using the reference method, Ref.\,~\cite{morreauPhononinducedDephasingQuantumdotcavity2019} (red) is compared with the optimized method (blue), and TEMPO (green). All calculations use the same number of neighbors, $L=26$, with the optimized and TEMPO methods using an SVD threshold of $\epsilon=10^{-8}$. Inset: Relative RMS error for the optimized approach (blue) and TEMPO (green) versus computational time across a range of SVD truncation thresholds. Error is calculated against reference values using $\epsilon=10^{-11}$ in each method. The phonon bath temperature is $T=50\,$K and the acoustic phonon parameters are taken as $D_c - D_v =-6.5\,$eV, where $D_c$ ($D_v$) is the conduction (valence) band deformation potential, $v_s=4.6\times 10^3\,$m/s is sound velocity, $\rho_m = 5.65\,\text{g}/\text{cm}^3$ is the mass density~\cite{muljarovDephasingQuantumDots2004,muljarovPhononInducedExcitonDephasing2005}.
}
\end{figure}

\Fig{QDCAVITY_Comparison} illustrates the accuracy of our optimization scheme by calculating the linear polarization $\mathrm{P}_{11}(t)$ of an isotropic (spherical) QD coupled to a microcavity with coupling strength $g=100\,\mu$eV and zero detuning ($\Omega_X=\Omega_C$).
% The parameters of InGaAs QDs studied in \cite{muljarovDephasingQuantumDots2004,muljarovPhononInducedExcitonDephasing2005} are used: $D_c - D_v =-6.5\,$eV, where $D_c$ ($D_v$) is the conduction (valence) band deformation potential, $v_s=4.6\times 10^3\,$m/s is sound velocity, $\rho_m = 5.65\,\text{g}/\text{cm}^3$ is the mass density, the confinement length $l=3.3\,$nm, and the temperature is $T=50\,$K.
We compare the results from the reference method \cite{morreauPhononinducedDephasingQuantumdotcavity2019}, the optimized calculation, and the TEMPO approach \cite{strathearnEfficientNonMarkovianQuantum2018}, using the same number of neighbors, $L=26$, in each case. The same SVD truncation threshold of $\epsilon =  10^{-8}$ is used in both the optimized and TEMPO calculations. The inset of \Fig{QDCAVITY_Comparison} shows the relative error, calculated via root mean square (RMS) of the polarization calculation found via the new optimized approach and TEMPO versus computational time for a range of SVD truncation thresholds. The ``ideal'' calculation used for the RMS error was a calculation using $\epsilon=10^{-11}$ for each technique. For the same SVD truncation threshold, the error is smaller in our approach but this is simply due to the technique performing less SVDs at each time step.
%The observed two orders of magnitude larger error using the TEMPO method is because this technique has more matrices and therefore performs more SVDs which accumulates the SVD truncation error. This is further discussed in \Sec{Extrapolation}.
There is also a benefit in computational time of our optimized approach. The computation takes approximately 1300 seconds using the reference method, compared with only 76 seconds using the optimization scheme for the same number of neighbors and $\epsilon=10^{-6}$. Notably, even decreasing the truncation threshold only slightly increases the computational time, taking just 85 seconds using $\epsilon=10^{-10}$. A calculation with TEMPO using $\epsilon =  10^{-6}$ took just 197 seconds, but $\epsilon =  10^{-10}$ took 2767 seconds, showing that the computational time of TEMPO depends more strongly on $\epsilon$. The different response to filtering via $\epsilon$ of our approaches clearly distinguishes the two techniques. Note also that TEMPO calculates more density matrix elements and so it is expected that computations take a few times longer. All computations were performed on an AMD 9800X3D processor with 32 GB DDR5 6000 MHz RAM. Furthermore, the memory storage requirement is over 4000 times less than in the reference method, for this number of neighbors. Note that the maximum achievable number of neighbors in the reference method for this QD-cavity system is $L=26$ on a desktop PC with 8GB RAM, whereas in the optimized case it is approximately $L=50$, which would require over 50 million GB of RAM using the reference method.

\section{Extrapolation}\label{Extrapolation}

As detailed in \cite{morreauPhononinducedDephasingQuantumdotcavity2019, hallControllingDephasingCoupled2025}, the strong coupling between the exciton and cavity forms a polariton, and the oscillatory behavior in \Fig{QDCAVITY_Comparison} is explained by the coherent exchange of quantum information between the QD and the cavity. The energy levels of the polariton states (hybrid QD-cavity states) are separated by the Rabi splitting, $R$, which determines the beat frequency in $|\mathrm{P}_{11}(t)|$.
The temporal decay of the linear polarization expresses the decoherence in this system as a consequence of the interaction of the QD with the bath. This decoherence is due to phonon-assisted transitions between the upper $\ket{+}$ and lower $\ket{-}$ polariton states, see \Sec{QDCAV_largeg} and \cite{hallControllingDephasingCoupled2025} for more details.

Using this picture, we have applied to the long-time dynamics of $\mathrm{P}_{11}(t)$ a biexponential fit of the form
\be
\label{Pfit}
\mathrm{P}_{11}^{\rm fit}(t)=\sum_j {C}_j e^{ -i\omega_j t }\,,
\ee
extracting the complex amplitudes ${C}_j$, energies $\Re\,\omega_j$, and dephasing rates $\Gamma_j=-\Im\, \omega_j$ of the phonon-dressed polariton states. In all of the single QD-cavity system calculations, we keep a fixed memory time and increase the number of neighbors $L$ to achieve a finer time discretization. It turns out that these extracted parameters follow a power law dependence on the number of neighbors $L$, for sufficiently large $L$.
Therefore, one can calculate the linear optical polarization for a given number of neighbors, apply a fit \Eq{Pfit} to the long time data, extract the fit parameters, and then apply a power-law fit to these parameters across an accessible range of $L$ values. The resulting parameters of this fit are used to perform a power-law extrapolation, estimating the exact value which corresponds to $L=\infty$.

As an example, consider the parameters corresponding to the dephasing rates, $\Gamma(L)$ extracted across a range of neighbors, where the convergence of $\Gamma(L)$ to the exact ($L=\infty$) value is assumed to follow a power law model, given by
\begin{equation}
	\Gamma(L) = \Gamma(\infty) + \alpha L^{-\beta}.
	\label{GamL}
\end{equation}
\begin{figure}[t]
	\centering
	\includegraphics[width=0.5\textwidth]{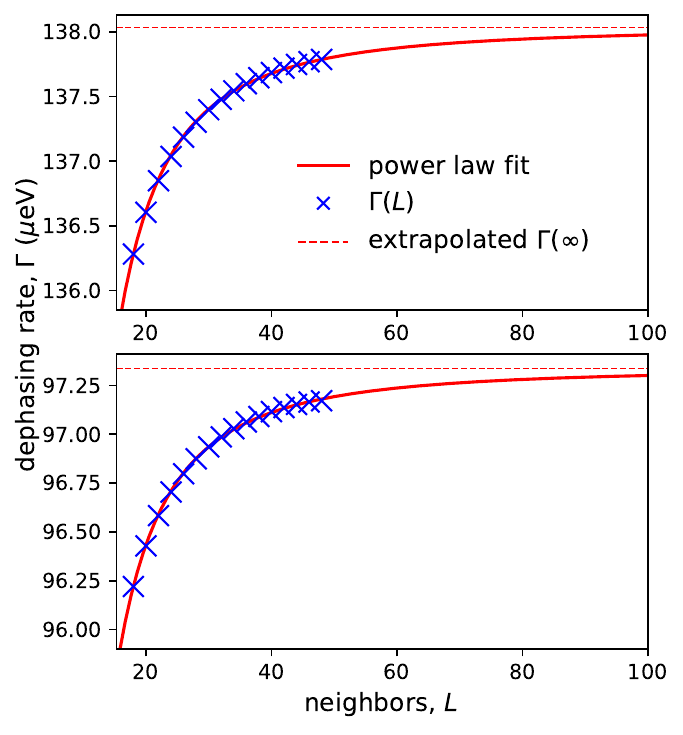}
	\caption{
		Power law fit applied to $\Gamma_+(L)$ (upper panel) and $\Gamma_-(L)$ (lower panel) across the range of $L$ values shown, for $g=600\,\mu$eV. The blue crosses are the values of $\Gamma_\pm(L)$ extracted from the biexponential fit, the red curves are the power law fit with $\beta=2$, and the red horizontal dashed lines are the extrapolated values of $\Gamma_\pm(\infty)$. The rest of the parameters are as in \Fig{QDCAVITY_Comparison}.
	}
	\label{PL_fit}	
\end{figure}
Figure~\ref{PL_fit} shows the $\Gamma(L)$ calculated values (blue crosses) in the QD-cavity system, with the power law model applied (red curve), and the extrapolated $\Gamma(\infty)$ is shown as a red dashed line. The value of $\Gamma(\infty)$ is estimated for the values of $\Gamma(L)$ shown in \Fig{PL_fit} by minimising the root mean square deviation from the power law \Eq{GamL} for $\beta=2$. Note that we first let $\beta$ be a fit parameter but found that it is always close to $\beta=2$, and therefore fixed this value. Physically, this behaviour can be attributed to the time discretization error of the Trotter decomposition, which scales as $\mathcal{O}(\Delta t^2)$. Since the time step $\Delta t$ scales as $1/L$, this implies that the resulting error scales as $1/L^2$, consistent with the observed power-law dependence of the extrapolated parameters.
\Fig{PL_fit} also demonstrates the necessity of the extrapolation, since optimization alone may not be well converged in some cases. However, the extra data provided by the optimization by achieving more neighbors allows for an accurate usage of the extrapolation. In fact, in some systems with very long memory times, such as in spatially extended quantum systems, the extrapolation is not possible at all with the reference method and requires the data from the optimization scheme. In TN approaches, parameter extrapolation is complicated by the interdependence of the SVD truncation threshold, $\epsilon$, and the number of neighbors $L$. In TN frameworks, the number of SVD operations scales with $L$, causing the truncation error to fluctuate across the calculation range. This prevents systematic extrapolation unless $\epsilon$ is sufficiently small to ensure that Trotterization remains the dominant error source. If this dominance is not maintained, the varying SVD error leads to non-monotonic convergence. We are presently unable to use the OQuPy/TEMPO package to generate a smooth convergence plot, as in \Fig{PL_fit}, even using very small SVD truncation values, e.g. $\epsilon =  10^{-8}$. Lowering this truncation threshold further increases computational time substantially and can take over a day for a single $L$ calculation, whereas simply choosing $\epsilon =  10^{-8}$ in our optimization across a few calculations with different $L$ and extrapolating them takes a matter of minutes. We verified the accuracy of this extrapolation by using the more memory-efficient TEMPO algorithm using $L=200$ neighbors. By taking this $L=200$ TEMPO value as exact, we find our extracted $\Gamma_\pm$ parameters have a relative difference of approximately $0.03\%$ or less, and only took several minutes.

\section{Results} \label{Results}

\begin{figure}[t]
	\centering
	\includegraphics[width=0.5\textwidth]{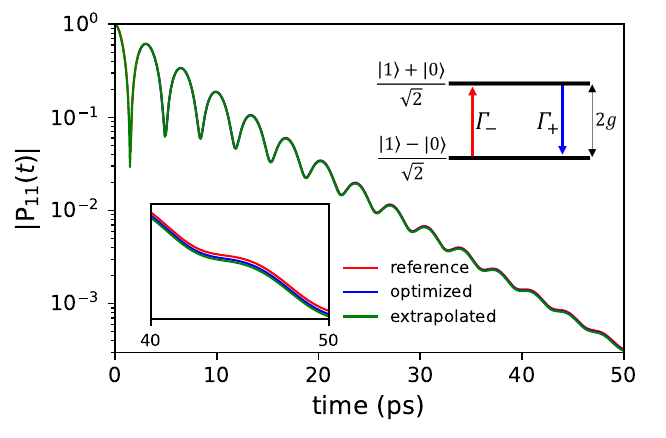}
	\caption{\label{QDCAVITY_g600}
		Linear optical polarization $|\mathrm{P}_{11}(t)|$ of an isotropic QD with $l=3.3\,$nm, coupled to a microcavity with coupling strength $g=600\,\mu$eV at zero detuning, calculated using the reference method with $L=26$ (red), the optimized scheme with $L=48$ and SVD threshold $\epsilon= 10^{-8}$ (blue), and extrapolation (green). Inset: Energy level diagram for the hybridized QD-cavity states $\ket{\pm}=(\ket{1}\pm\ket{0})/\sqrt{2}$, with real phonon-assisted transitions (red and blue arrows). The phonon bath parameters are as in \Fig{QDCAVITY_Comparison}.
}
\end{figure}

\subsection{Accessing stronger coupling regimes in the QD-cavity system}\label{QDCAV_largeg}
\Fig{QDCAVITY_g600} shows the linear optical polarization $\mathrm{P}_{11}(t)$ of an isotropic QD coupled to a microcavity with coupling strength $g=600\,\mu$eV at zero detuning.
The data is calculated with the reference method, the optimization and optimization including extrapolation.
There is good convergence of the reference method calculation at this coupling strength because the time step $\Delta t$, which decreases as the number of neighbors increases, is small enough to resolve the dynamics of the Rabi oscillations. The frequency of the Rabi oscillations is determined by the energy splitting between the polariton states, which is $R\approx 2g$ at zero detuning (see the right inset of \Fig{QDCAVITY_g600}). Therefore, as $g$ increases, a smaller time step is required for an accurate calculation. The optimization and extrapolation thus provides access to larger coupling strengths.

\begin{figure}[t]
	\includegraphics[width=0.5\textwidth]{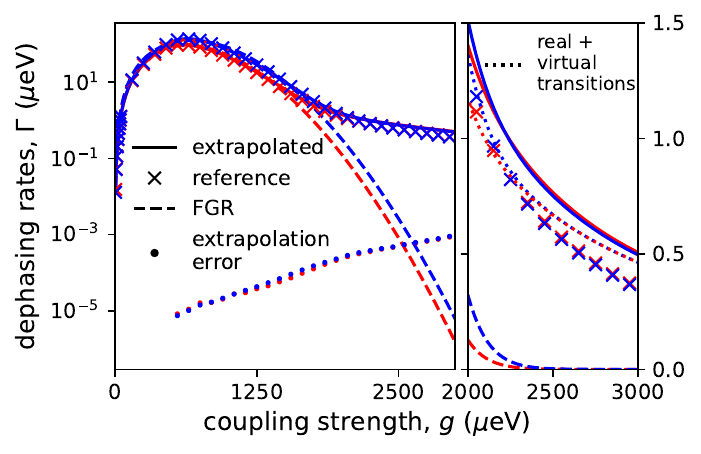}
	\caption{\label{QDCAVITY_vs_g}
		Dephasing rates $\Gamma_+$ (blue) and $\Gamma_-$ (red) of the hybridized states $|\pm\rangle$ at zero detuning as functions of the QD-cavity coupling strength $g$, calculated via the optimization scheme with extrapolation (solid lines), via FGR (dashed lines), and the most accurate calculation available via the reference method (crosses) for isotropic QDs with $l=3.3\,$nm. The error of the extrapolation (dots) is shown on the left subplot. The right subplot further includes the virtual transitions on top of the real ones (dots). The phonon bath parameters are as in \Fig{QDCAVITY_Comparison}.
}
\end{figure}

\Fig{QDCAVITY_vs_g} shows the dephasing rates $\Gamma_\pm$ in the QD-cavity system as a function of the coupling strength $g$. The optimized and extrapolated data visually align with the reference method calculation at smaller coupling strengths, but show a deviation at larger values as expected due to lack of convergence. Provided that the parameters extracted via the optimization scheme follow the power law convergence, the extrapolated values are much more accurate for significantly larger coupling strengths than in the reference method, Ref.\, \cite{morreauPhononinducedDephasingQuantumdotcavity2019}. The inset of \Fig{QDCAVITY_vs_g} shows the extrapolated error for each of the extrapolated $\Gamma_\pm$ values, where even at $g=3000\,\mu$eV the extrapolated error ($10^{-3} \, \mu$eV) is much less than the difference between the extrapolated and reference method calculations ($10^{-1}\, \mu$eV). This error is estimated by taking the final 4 data points of $\Gamma_{\pm}(L)$, performing the extrapolation with the last 3 points, and comparing the result with the extrapolation obtained using the first and last 2 points.

It is worth noting that the discrepancy between Fermi's golden rule (FGR) and the full calculation at large coupling strengths is due to virtual phonon-assisted transitions, not accounted for by the standard FGR calculation of real phonon-assisted transitions. In fact, in the regime of large coupling strengths, often referred to in the literature as the dynamical decoupling regime \cite{denning_optical_2020}, virtual transitions become comparable to, or even dominate over, real phonon-assisted processes \cite{muljarovDephasingQuantumDots2004, muljarovPhononInducedExcitonDephasing2005}. While, in principle, multi-phonon processes could bridge the energy gap between the upper and lower states, such real phonon-assisted transitions must satisfy energy conservation, rendering their probability negligibly small. In contrast, virtual transitions are not constrained by energy conservation and can therefore dominate at large coupling strengths. Within our approach, the cumulant expansion naturally includes such virtual processes to all orders, allowing us to capture multi-phonon effects. On the analytical side, the FGR can be refined by incorporating virtual transitions, for example by using the quadratic coupling model \cite{muljarovDephasingQuantumDots2004}, or by employing a combined approach \cite{zibikIntersublevelPolaronDephasing2008}. In this work, we adopt the quadratic coupling model \cite{muljarovDephasingQuantumDots2004}  to include both real and virtual phonon-assisted transitions up to second order in the cumulant. Details on FGR for real single phonon-assisted transitions and the refined model incorporating virtual transitions are provided in \App{App:FGR}.
In \Fig{QDCAVITY_vs_g}, one can see that at large coupling strengths $g$ (and thus large Rabi splittings $R$), the dephasing rates calculated via FGR for single real phonon-assisted transitions are exponentially small, approaches zero, despite the full calculation yielding finite values, whereas the modified calculation including virtual transitions shows a good agreement with the full extrapolated calculation. This agreement supports our conclusion on the origin of the decoherence in this regime and serves as a verification of the extrapolated results. The calculation via the reference method at such large coupling strengths begins to deviate significantly from both the optimized and extrapolated calculation. Thus, the optimization scheme combined with extrapolation allowed us to demonstrate the physical mechanism of decoherence in a QD-cavity system in very strong coupling regime, which is understood and quantified in terms of virtual phonon-assisted transitions.

%{\bf EM: Please add to the figure and give an equation for the combination of real and virtual with some details, such as strict definitions of $\Gamma_\pm(\omega)$ functions in a new appendix, say in words here what do they mean. Be careful to distinguish them from $\Gamma_\pm(L)$ used above -- need different notations.}
%
%{\bf EM: Add here a conclusion on the extrapolation: The agreement of the real+virtual transitions model with the extrapolated values confirms supports the validity of the refinement used. It seems they are closer to the extrapolated than to the original data, right? Also add that the optimization combined with the extrapolation allowed us to demonstrate the physical mechanism (virtual transitions).}

\subsection{Necessity of the optimization in spatially extended systems}
When the parts of a spatially extended quantum system interacts with the same environment, the memory time of the system can become very large and its quantum dynamics is difficult to capture accurately.
As a typical example of this, we now consider {\em Case 2}, a system of two spatially separated QDs coupled to a microcavity and interacting with a shared phonon bath.
Physically, the long memory times in this case are due to the shared bath, with coherent phonons travelling between the QDs, so the larger the dot separation, the larger the memory time. That is, in these spatially extended systems, the memory time is no longer constant as in single QD systems and depends on the distance between the QDs.

\begin{figure}[t]
	\includegraphics[width=0.5\textwidth]{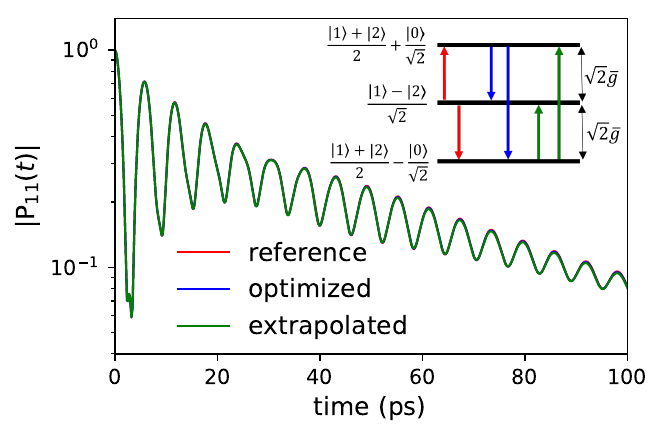}
	\caption{\label{QDQDCAVITY_d5}
		Linear polarization $|\mathrm{P}_{11}(t)|$ for cavity-mediated ($g_1=g_2=\bar{g}=500\mu\,$eV) isotropic QDs with $l=3.3\,$nm, at zero detuning ($\Omega_1=\Omega_2=\Omega_C$) and QDs separation $d=5\,$nm, with excitation and measurement in QD $1$, calculated via the reference method with $L=16$ (red),  optimization with $L=30$ and SVD threshold $\epsilon= 10^{-8}$ (blue), and extrapolation (green). Inset: Energy level diagram for the hybridized QD-cavity states $|+,\pm\rangle=(|1\rangle+|2\rangle)/2\pm|0\rangle/\sqrt{2}$ and $|-\rangle=(|1\rangle-|2\rangle)/\sqrt{2}$, with real phonon-assisted transitions (red, blue and green arrows). The phonon bath is at $T=20\,$K, and other parameters are as in \Fig{QDCAVITY_Comparison}.
}
\end{figure}

\Fig{QDQDCAVITY_d5} shows the linear polarization, $|\mathrm{P}_{11}(t)|$, for a pair of cavity-mediated ($g_1=g_2=\bar{g}=500\mu\,$eV) coupled isotropic QDs separated by a center-to-center distance $d=5\,$nm. In the case of zero detuning ($\Omega_1=\Omega_2=\Omega_C$) and equal QD-cavity couplings ($g_1=g_2$), there are three resulting hybridized states, $|+,\pm\rangle=(|1\rangle+|2\rangle)/2\pm|0\rangle/\sqrt{2}$ and  $|-\rangle=(|1\rangle-|2\rangle)/\sqrt{2}$, as depicted in the inset. Here $|1\rangle$ and $|2\rangle$ denote the exciton states in QD 1 and 2, respectively, and $|0\rangle$ is the cavity-photon state, see \App{App:Trotter-Linked} and \cite{hallControllingDephasingCoupled2025} for details. To generate the extrapolated data, a triexponential fit \Eq{Pfit} is applied to $P_{11}(t)$ to extract the parameters corresponding to these hybridized states at different $L$ and then extrapolate them to $L\to \infty$ using a power law fit similar to \Eq{GamL} and  the procedure described in \Sec{Extrapolation}. The full extrapolated linear polarization is then found for any observation time $t$ by adding $\Delta P_{11}(t)=P_{11}(t)-P_{11}^{\rm fit}(t)$ to the triexponential form \Eq{Pfit} used for the extrapolated parameters $C_j(\infty)$ and $\omega_j(\infty)$. Owing to the finite memory time, the difference $\Delta P_{11}(t)$ is independent of $L$, at least for a sufficiently large achievable $L$, and is found by subtracting from the full linear polarization $P_{11}(t)$, calculated at a finite $L$, its long-time dependence \Eq{Pfit} with the extracted fit parameters $C_j(L)$ and $\omega_j(L)$.
\Fig{QDQDCAVITY_d5} compares the reference method, optimized, and extrapolated calculations, demonstrating a visual agreement between them. However, the separation of $5\,$nm is rather small and does not significantly increase the memory time, as detailed in Ref.\,\cite{hallControllingDephasingCoupled2025}. We therefore observe a sufficient convergence of the reference method calculation, as in the QD-cavity system with small or moderate coupling $g$, see \Figs{QDCAVITY_Comparison}{QDCAVITY_g600}.

\begin{figure}[t]
	\includegraphics[width=0.5\textwidth]{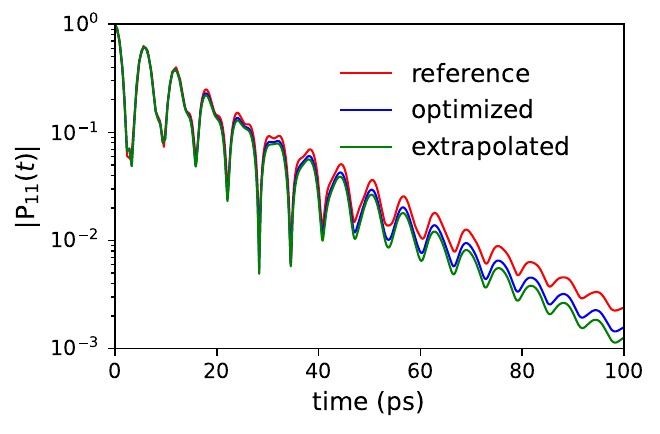}
	\caption{\label{QDQDCAVITY_d45}
		As \Fig{QDQDCAVITY_d5} but with the QD separation of $d=45\,$nm. }
\end{figure}

\Fig{QDQDCAVITY_d45} shows the same quantity, but instead with a QD separation of $d=45\,$nm, which displays a clear lack of convergence using the reference method. Even the calculation using the optimization scheme is not very well converged, showing the necessity of the extrapolation. It should be noted that the extracted parameters do not follow a power-law convergence across all $L$ values. Instead, the selected $L$ values must be sufficiently high to enter the regime of power-law convergence. Due to this, the reference method in most regimes of spatially extended systems cannot generate enough data within the power-law convergence regime to provide an extrapolated calculation. Therefore, the optimized calculations are required for the extrapolation.
\begin{figure}[t]
	\centering
	\includegraphics[width=0.5\textwidth]{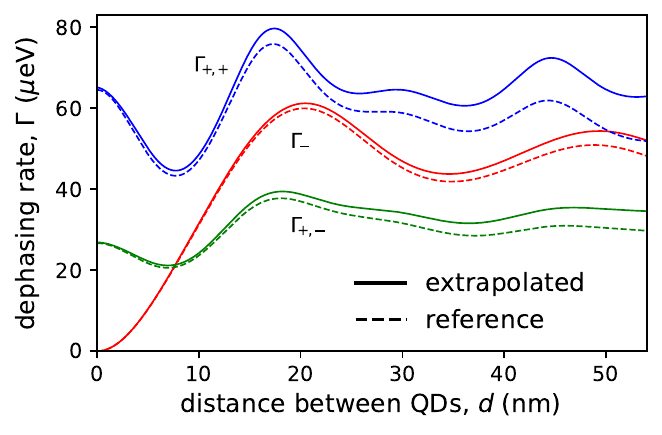}
	\caption{Dephasing rates $\Gamma_{+,\pm}$ and $ \Gamma_{-}$ of the hybridized states as functions of $d$. Extrapolated rates (solid lines) are compared with rates found via the reference method (dashed lines), for cavity-mediated coupled isotropic QDs with $l=3.3\,$nm, interaction strength $g_1=g_2=\bar{g}=500\,\mu$eV, and zero detuning. The phonon bath parameters are as in \Fig{QDQDCAVITY_d5}.}
	\label{QDQDCAV_Dephasing}
\end{figure}

\Fig{QDQDCAV_Dephasing} shows the dephasing rates $\Gamma_j$ extracted from the fit as functions of the QD separation $d$. The dephasing rates of the hybridized states $|+,\pm\rangle$ and  $|-\rangle$ are denoted by $\Gamma_{+,\pm}$ and $\Gamma_{-}$, respectively. As the QD separation increases, the memory time increases, and the reference method calculation gradually becomes increasingly inaccurate, requiring the optimization and extrapolation to accurately model the system.

\section{Conclusion}
In conclusion, we have developed a general optimization for path-integral tensor-multiplication schemes used to compute the exact non-Markovian dynamics in a large class of open quantum systems, with pair correlations and finite memory times. We demonstrate the optimization by investigating the linear polarization in quantum dot-cavity systems, and verify the scheme by comparison with the reference method \cite{morreauPhononinducedDephasingQuantumdotcavity2019} and the TEMPO algorithm \cite{fuxOQuPyPythonPackage2024}. The optimized scheme has largely reduced memory storage requirements and approximately doubles the number of neighboring connections within the tensor network structure.
Unlike other approaches, it provides a reliable dependence of the calculated quantum dynamics on the time step $\Delta t$ of discretization, that allows us to reach Trotter's limit $\Delta t\to0$ by extrapolation. This is crucial for systems with weak convergence to the exact solution.
The increased number of neighbors paired with extrapolation gives access to previously inaccessible regimes, offering better convergence for systems with long memory times.
We have considered two examples of such weak convergence in which the extrapolation allowed us to accurately calculate the quantum dynamics, revealing mechanisms of decoherence: QD-cavity in very strong coupling regime and cavity-mediated coupled qubits at large distances. For calculations already well converged using traditional approaches, the optimization scheme offers greatly improved computational efficiency, shortening calculation times by up to two orders of magnitude.

\section*{Acknowledgments}
L.H. acknowledges support from the EPSRC under
grant no. EP/T517951/1.

\section*{Data Availability}
The data that support the findings of this article are openly available \cite{Hall2026Data}.

\appendix

\section{System Hamiltonians}
\label{SysHam}
In this work, we apply the optimization to two systems, referred to in the main text as {\em Case 1}: a QD-cavity system coupled to a bath of acoustic phonons, as detailed in \cite{morreauPhononinducedDephasingQuantumdotcavity2019}, and {\em Case 2}: a QD-QD-cavity system coupled to the same acoustic phonon bath, as detailed in \cite{hallControllingDephasingCoupled2025}.

In {\em Case 1}, the full Hamiltonian has the form
\begin{equation}
	H = H_{0} + H_\text{\rm IB}\,,
	\label{Ham_case1}
\end{equation}
where
\begin{equation}\label{H0_case1}
		H_{0} = \Omega_X d^\dagger d + \Omega_C a^\dagger a + g (d^\dagger a + a^\dagger d)
\end{equation}
describes the coupling between the QD and the cavity, and $H_\text{\rm IB}$ is an independent boson (IB) model Hamiltonian describing the coupling of the QD to the environment. Here, $d^\dagger$ is the fermionic exciton creation operator in the QD, $a^\dagger$ is the cavity photon creation operator, $\Omega_{X}$ ($\Omega_{C}$) is the exciton (cavity photon) energy, $g$ is the coupling strengths between the QD and cavity. The IB model Hamiltonian describes the interaction of the QD exciton with an acoustic-phonon bath,
\begin{equation}\label{HIB_case1}
	H_\text{\rm IB} = H_\text{\rm B} + H_{\mathrm{int}} \,,
\end{equation}
where
\begin{equation}\label{Hph_V_case1}
	H_\text{\rm B}=\sum_{\textbf{q}} \omega_q b_{\textbf{q}}^\dagger b_{\textbf{q}}\ \ \ {\rm and} \ \ \
	H_{\mathrm{int}} = d^\dagger d\sum_{\textbf{q}} \lambda_{\textbf{q}} (b_{\textbf{q}} + b_{-\textbf{q}}^\dagger)
\end{equation}
are, respectively, the free phonon bath Hamiltonian and the QD coupling to the bath, where $b_{\textbf{q}}^\dagger$ is the bosonic creation operator of a bulk phonon mode with the momentum $\textbf{q}$ and frequency $\omega_q$ (denoting $q = \lvert \textbf{q} \rvert$). The coupling of the exciton in the QD to the phonon mode $\textbf{q}$ is given by the matrix element $\lambda_{\textbf{q}}$,
which depends on the material parameters and the QD exciton wave function. Its explicit form for isotropic QDs is provided in \App{App:Coupling}.

%The explicit Hamiltonian for {\em Case 2} is:
%\begin{equation}
%	H = H_{0} + H_\text{\rm IB}\,,
%	\label{Ham_case2}
%\end{equation}
%where $H_0$ describes the coupling between the qubits and the cavity, and $H_\text{\rm IB}$ is a generalized independent boson (IB) model Hamiltonian describing the coupling of the qubits to the shared environment.
In {\em Case 2}, for a system of two remote QDs coupled to an optical cavity, $H_0$ takes the form
\begin{equation}\label{H0_case2}
	\begin{split}
		H_{0} =& \; \Omega_1 d_1^\dagger d_1 + \Omega_2 d_2^\dagger d_2 + \Omega_C a^\dagger a \\ &+ g_1 (d_1^\dagger a + a^\dagger d_1) + g_2 (d_2^\dagger a + a^\dagger d_2)\,,
	\end{split}
\end{equation}
where  $d_{j}^\dagger$ is the fermionic exciton creation operator in QD $j$ ($j=1,2$), $a^\dagger$ is the cavity photon creation operator, $\Omega_{j}$ ($\Omega_{C}$) is the exciton (cavity photon) frequency, and $g_{j}$ is the coupling strength between the exciton in QD $j$ and the cavity photon. The generalized IB model Hamiltonian describes the interaction of the QD excitons with a shared acoustic-phonon bath,
\begin{equation}\label{HIB_case2}
	H_\text{\rm IB} = H_\text{\rm B} + H_{\mathrm{int}}\,,
\end{equation}
where $H_\text{\rm B}$ is given in \Eq{Hph_V_case1} and
\begin{equation}\label{Hph_V_case2}
		H_{\mathrm{int}} = \sum_{j=1,2} d_j^\dagger d_j \sum_{\textbf{q}} \lambda_{\textbf{q},j} (b_{\textbf{q}} + b_{-\textbf{q}}^\dagger)
\end{equation}
is the coupling of the QD excitons to the bath, with $b_{\textbf{q}}^\dagger$, $\textbf{q}$, and $\omega_q$ defined above. It is convenient to write the interaction as
\be
H_{\mathrm{int}}= d_1^\dagger d_1 V_1+  d_2^\dagger d_2 V_2\,,
\ee
where
\be
V_j=\sum_{\textbf{q}} \lambda_{\textbf{q},j} (b_{\textbf{q}} + b_{-\textbf{q}}^\dagger)\,.
\label{V_case2}
\ee
The coupling of the exciton in QD $j$ to the phonon mode $\textbf{q}$ is given by the matrix element $\lambda_{\textbf{q},j}$ with its explicit form for isotropic QDs provided in \App{App:Coupling}. For identical QDs separated by a distance vector $\textbf{d}$, the matrix elements satisfy the relation
\begin{equation}\label{lambda_case2}
	\lambda_{\textbf{q},2} = e^{i\textbf{q}\cdot \textbf{d}}\lambda_{\textbf{q},1}\,.
\end{equation}

\section{Exciton-phonon coupling elements and phonon spectral density}
\label{App:Coupling}

Throughout this work, we use typical InGaAs QD parameters \cite{muljarovDephasingQuantumDots2004,muljarovPhononInducedExcitonDephasing2005}, which are listed in the caption to \Fig{QDCAVITY_Comparison}. At low temperatures, the exciton-phonon interaction is primarily governed by the deformation potential coupling with longitudinal acoustic phonons. Assuming that the phonon parameters within the QDs closely resemble those of the surrounding material, and further assuming that the acoustic phonons exhibit linear dispersion, $\omega_q = v_s q $, where $q=|{\bf q}|$ and $v_s$ is the sound velocity in the material, the exciton-phonon coupling matrix element for an exciton in QD $j=1,2$ is given by
\begin{equation} \label{lambda_D}
	\lambda_{\textbf{q},j} = \frac{\sqrt{q}\,\mathcal{D}_j(\textbf{q})}{\sqrt{2\rho_m v_s \mathcal{V}}}\,,
\end{equation}
where $\rho_m$ is the mass density of the material, $\mathcal{V}$ is the sample volume, and
\begin{equation}\label{form-factor}
	\mathcal{D}_j(\textbf{q})= \int d\textbf{r}_{e} \int d\textbf{r}_{h}  \abs{\Psi_{X,j}(\textbf{r}_{e},\textbf{r}_{h})}^2 \left(D_c e^{i\textbf{q}\cdot \textbf{r}_{e}} -D_v e^{i \textbf{q}\cdot \textbf{r}_{h}} \right)
\end{equation}
is the coupling form-factor \cite{muljarovDephasingQuantumDots2004, muljarovPhononInducedExcitonDephasing2005}, with $D_{c}$ ($D_{v}$) being the deformation potential of the conduction (valence) band. Assuming a factorizable form of the exciton wave functions, $\Psi_{X,j}(\textbf{r}_{e},\textbf{r}_{h})= \psi_{e,j}(\textbf{r}_{e}) \psi_{h,j}(\textbf{r}_{h})$, where $\psi_{e(h),j}(\textbf{r})$ is the confined electron (hole) ground state wave function in QD $j$, the form-factor simplifies to
\begin{equation}\label{deformationpotential}
	\mathcal{D}_j(\textbf{q}) = \int d\textbf{r} \left[ D_c\abs{\psi_{e,j}(\textbf{r})}^2 - D_v \abs{\psi_{h,j}(\textbf{r})}^2\right] e^{i\textbf{q} \cdot \textbf{r} }\,.
\end{equation}

Choosing spherically symmetric parabolic confinement potentials, the ground-state wave functions of the carriers take Gaussian form, which in the simpler case of equal electron and hole confinement lengths, $l_{e,j} = l_{h,j} = l_j$, is given by
\begin{equation} \label{psi_j}
	\psi_{e,j}(\textbf{r}) =\psi_{h,j}(\textbf{r}) = \frac{1}{\pi^{3/4} l_j^{3/2}} \exp{-\frac{ (\textbf{r} - \textbf{d}_j)^2}{2l_j^2}}\,,
\end{equation}
where $\textbf{d}_j$ is the coordinate of the center of QD $j$.  Substituting \Eq{psi_j} into \Eq{deformationpotential}, performing the integration over the whole space and using the result in \Eq{lambda_D}, we obtain
\begin{equation}\label{lambda_j}
	\lambda_{\textbf{q},j} = \sqrt{q} \lambda_0 \exp{-l_j^2 q^2/4} e^{i \textbf{q}\cdot\textbf{d}_j}\,,
\end{equation}
where
\begin{equation}
	\lambda_0 = \frac{D_c - D_v}{\sqrt{2\rho_m v_s \mathcal{V}}}\,.
\end{equation}
Choosing the first QD located at the origin ($\textbf{d}_1=0$) we find $\textbf{d}_2=\textbf{d}$, where $\textbf{d}$ is the distance vector between the QDs.
Converting the summation over ${\bf q}$ to an integration, $\sum_{\bf q} \rightarrow \frac{\mathcal{V}}{(2\pi)^3} \int d {\bf q}$, and using spherical coordinates, we find the phonon spectral density
\begin{equation}\label{spectraldens_isotropic}
	J_{jj'}(\omega)= {\it J} \, \omega^3 \exp{-\frac{ \omega^2 l^2}{v_s ^2}}
	\times
	\begin{cases}
		1 & j=j' \\
		{\rm sinc}\left(\frac{\omega d}{v_s}\right) & j \neq j'\,,
	\end{cases}
\end{equation}
where
\begin{equation}\label{J0}
	{\it J} = \frac{(D_c - D_v)^2}{4\pi^2 \rho_m v_s^5}\,,
\end{equation}
$d=|{\bf d}|$, $l^2 = (l_j^2 + l_{j'}^2)/4$ (for brevity omitting the indices $j$ and $j'$ in the new length $l$ introduced), and ${\rm sinc}\,(x)=\sin x/x$.
However, in {\em Case 1} only the $j=j'$ case is required.

The spectral density $J_{jj'}(\omega)$ is used in \App{App:Trotter-Linked} below for calculating the IB model cumulants.

%\section{Linear polarization}
%\label{Method}
%In this work, we focus on the linear optical polarization, which is the simplest correlation function, which allows us in particular to study the coherence of the system in {\em Case 2} as a function of the distance between the QDs. The linear polarization of QD $j$ is defined as
%$P_{jk}(t)={\rm Tr}\{\rho(t) d_j\}$, where $\rho(t)$ is the full density matrix and the index $k$ denotes the QD which is instantaneously excited at time $t=0$. As has been derived in Ref.\,\cite{morreauPhononinducedDephasingQuantumdotcavity2019}, the linear polarization can written as
%\begin{equation}\label{LinPol}
%	P_{jk}(t) = \langle \bra{j} \hat{U}(t) \ket{k} \rangle_\text{\rm B}\,,
%\end{equation}
%where $\hat{U}(t) = e^{iH_\text{\rm B}t} e^{-iHt}$ is the evolution operator and $\langle...\rangle_\text{\rm B}$ denotes the expectation value over all phonon degrees of freedom in thermal equilibrium. Here and below we use the following basis states
%\begin{equation}
%	\ket{j}=d_j^\dagger \ket{0} \rmand \ket{C}= a^\dagger \ket{0}\,,
%\end{equation}
%where $\ket{0} $ represents the vacuum state of the QD-cavity subsystem, and $j=1,\,2$.
%
%Taking advantage of the two exactly solvable parts of the Hamiltonian, $H_0$ and $H_{\mathrm{IB}}$, we apply the rigorous method of Trotter's decomposition with cumulant expansion~\cite{morreauPhononinducedDephasingQuantumdotcavity2019}, which we summarize in the section below, that allows us to take into account the effect of the phonon environment {\it exactly}.

\section{Trotter's decomposition with linked cluster expansion}
\label{App:Trotter-Linked}
\subsection{Trotter's decomposition}

The time evolution operator \Eq{TrottersU_mt} can be written in terms of two operators $\hat{M}$ and $\hat{W}$, which describe the dynamics due to $H_0$ and $H_\text{\rm IB}$, respectively, each part being analytically solvable. Using these operators, the dynamics without phonons over a single time step is described by $\hat{M}$, given by \Eq{Moperator}, and the exciton-phonon dynamics is given by
\begin{equation}\label{WoperatorForster}
	\hat{W}(t_n,t_{n-1}) = e^{iH_\text{\rm B}t_n}e^{-iH_\text{\rm IB}\Delta t} e^{-iH_\text{\rm B}t_{n-1}}\,,
\end{equation}
where $t_n = n\Delta t$. Exploiting the commutivity of $H_0$ and $H_\text{\rm B}$, one can write the time evolution operator \Eq{TrottersU_mt} as
\begin{equation}\label{UasproductWM}
	\hat{U}(t) = \mathcal{T} \prod_{n=1}^N \hat{W}(t_n,t_{n-1})\hat{M}(t_n -t_{n-1})\,,
\end{equation}
where $\mathcal{T}$ is the time-ordering operator. Focusing on {\it Case 2}, we use the following basis states
\begin{equation}
\ket{0}= a^\dagger \ket{\rm VS}	\rmand \ket{j}=d_j^\dagger \ket{\rm VS}\,,
\end{equation}
where $\ket{\rm VS} $ represents the vacuum state of the QD-QD-cavity subsystem, and $j=1,\,2$.
$\hat{W}$ and $\hat{M}$ are both $3 \times 3$ matrices in the $\ket{0}$, $\ket{1}$, $\ket{2}$ basis, and due to the diagonal form of the exciton-phonon interaction, $\hat{W}$ is diagonal. Its diagonal matrix elements can be written as
\begin{equation}
	W_{i_n} (t_n,t_{n-1}) =  \mathcal{T} \exp{ -i \int_{t_{n-1}}^{t_n} \tilde{V}_{i_n}(\tau) d\tau }\,,
\end{equation}
where
\begin{equation}\label{Vbetanu}
	\tilde{V}_{i_n}(\tau) = \xi_{i_n} {V}_1(\tau) + \eta_{i_n} {V}_2(\tau)
\end{equation}
for $\tau$ within the time interval $t_{n-1} \leqslant \tau \leqslant t_{n}$\,, with  $\xi_{i}$ and $\eta_{i}$ being the components of the vectors
\be
\label{xi-eta}
\vec{\xi} = \begin{pmatrix} 0\\1 \\0 \end{pmatrix} \rmand \vec{\eta} = \begin{pmatrix} 0\\0 \\1 \end{pmatrix}\,,
\ee
respectively, and $V_j(\tau) = e^{iH_\text{\rm B}\tau}V_je^{-iH_\text{\rm B}\tau}$ is the exciton-phonon coupling in the interaction representation, with $V_j$ given by \Eq{V_case2}. We use the indices $i_n$ to indicate which state the system is in at a given time step $n$, being either $\ket{0}$, $\ket{1}$, or $\ket{2}$, with $i_n$ taking the values 0, 1, or 2, respectively. The elements of the vectors $\vec{\xi}$ and $\vec{\eta}$ take care of the choice for which exciton is coupled to phonons at time step $n$. For example, if the system is in the first QD exciton state at time step $n$, then $i_n =1$, and the exciton-phonon interaction $V_1$ occurs. Similarly, if $i_n = 0$, then both $\xi$ and $\eta$ are zero, since no exciton-phonon interaction occurs when the system is in the cavity state.
%{\bf EM: I would give details (the vectors and states) for Case 1 also, and note that this is a general approach }.

To find the polarization, we substitute the time-evolution operator $\hat{U}(t)$ in terms of the time ordered products of matrices $\hat{W}$ and $\hat{M}$ into \Eq{LinPol_mt}. Then writing the matrix products explicitly yields
\bea
P_{jk}(t) &=& \sum_{i_{N-1}} \dots \sum_{i_{1}} M_{i_N i_{N-1}}
\dots M_{i_1 i_{0}}
\nonumber\\
&&\times\langle W_{i_N}(t,t_{N-1}) \dots W_{i_1}(t_1,0)\rangle_\text{\rm B}\,,
\label{PolarizationMthenWForster}
\eea
where $i_0 = k$ and $i_N = j$ denote, respectively, the excitation mode $k$ at $t=0$ and measurement mode $j$ is at the final time step $t_N=t$, and $M_{i_n i_m}=[\hat{M}(\Delta t)]_{i_n i_m}$. The  $W_{i_n}$ operators fully include the phonon contributions, therefore we separate this product from the rest of the expression in order to take the expectation value and apply the linked cluster theorem  \cite{mahanManyParticlePhysics2000,muljarovDephasingQuantumDots2004,muljarovPhononInducedExcitonDephasing2005}.

\subsection{ Linked cluster expansion}\label{LCE}

To calculate the expectation value of the products of the exciton-phonon interaction operators in \Eq{PolarizationMthenWForster}, we apply the linked cluster theorem. It allows us to write this expectation value as an exponential with a double sum over all possible second-order cumulants in the exponent \cite{mahanManyParticlePhysics2000,morreauPhononinducedDephasingQuantumdotcavity2019},
\begin{equation}\label{tracecumulantForster}
	\langle W_{i_N}\!(t,t_{N-1}\!)\!\dots \! W_{i_1}\!(t_1,0)\rangle_{\rm B} \!=\! \exp(\sum_{n=1}^N \!\sum_{m=1}^N  \!\mathcal{K}_{i_n i_m}(|n\!-\!m|)\!\!)
\end{equation}
with the cumulants in \Eq{tracecumulantForster} given by
\begin{equation}\label{Knm}
	\mathcal{K}_{i_n i_m}(s)  =   -\frac{1}{2} \int_{t_{n-1}}^{t_n}d\tau_1 \int_{t_{m-1}}^{t_m} d\tau_2
	\langle \mathcal{T} \tilde{V}_{i_n}(\tau_1) \tilde{V}_{i_m}(\tau_2)\rangle_\text{\rm B},
\end{equation}
where $s=|n-m|$ and the double integration is performed over the time intervals $t_n - t_{n-1}$ and $t_m - t_{m-1}$.
Using \Eq{Vbetanu}, this cumulant can be expressed as
\bea
\mathcal{K}_{i_n i_m}(s) &=&  \xi_{i_n}\xi_{i_m} K_{11}(s) + \eta_{i_n} \eta_{i_m} K_{22}(s)
\nonumber\\
&&+ (\xi_{i_n}\eta_{i_m}+\eta_{i_n}\xi_{i_m})K_{12}(s)\,,
\label{Kinim}
\eea
where we have introduced the cumulant elements
\begin{equation}
	\label{Kjj'}
	K_{jj'}(s) =  -\frac{1}{2}  \int_{t_{n-1}}^{t_n} d\tau_1 \int_{t_{m-1}}^{t_m} d\tau_2 D_{jj'} (\tau_1 - \tau_2)\,,
\end{equation}
possessing the symmetry $K_{jj'}(s)=K_{j'j}(s)$. Here
\begin{equation}
	D_{jj'}(t) = \int_0^\infty d\omega \; J_{jj'}(\omega) D(\omega,t)\,,
\end{equation}
with
\begin{equation}\label{spectraldens}
	J_{jj'}(\omega) = \sum_\textbf{q} \lambda_{\textbf{q},j}\lambda_{\textbf{q},j'}^\ast \delta (\omega-\omega_q)\,,
\end{equation}
and
\begin{equation}
	D(\omega,t) = ({\cal N}_\omega +1) e^{-i\omega |t| }  + {\cal N}_\omega  e^{i\omega |t|}\,
\end{equation}
being, respectively the phonon spectral density and the standard phonon propagator, in which  ${\cal N}_\omega=[e^{\omega/(k_B T)}-1]^{-1}$ is the Bose distribution function, where $T$ is the temperature, and $k_B$ is the Boltzmann constant. The explicit analytical form of the spectral density $J_{jj'}(\omega)$ for isotropic QDs are derived in \App{App:Coupling}.

A convenient way to compute the cumulant elements \Eq{Kjj'} is by defining the cumulant function
\bea
\label{Cjj'}
C_{jj'}(t)&=& -\frac{1}{2}  \int_{0}^{t} d\tau_1 \int_{0}^{t} d\tau_2 \; D_{jj'}(\tau_1 - \tau_2)\, .
\eea
\begin{figure}[t]
	\centering
	\includegraphics[scale=0.65]{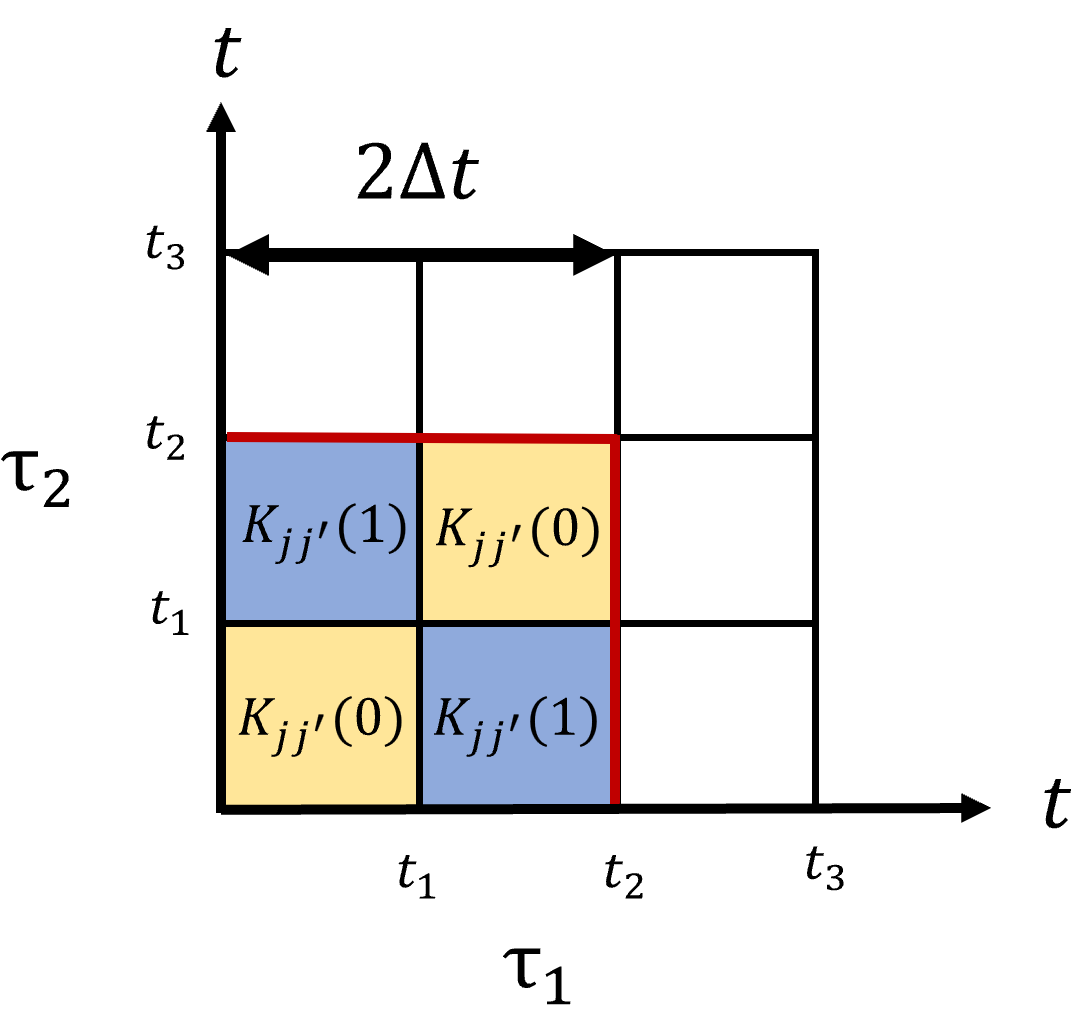}% Here is how to import EPS art
	\caption{\label{Kjj'_compute} A portion of the time grid up to $t=2\Delta t$, illustrating \Eq{cumfig} that $C_{jj'}(2\Delta t)$ is composed of diagonal $K_{jj'}(s=0)$ (yellow) and off-diagonal contribution $K_{jj'} (s=1)$ (blue). }
\end{figure}
Using the fact that $K_{jj'}(s)$ depends on the difference $|n-m|$ only, and not on both time steps $t_n$ and $t_m$ individually every possible cumulant element across any time step can be computed recursively via
\begin{align}
	\label{recursiveK}
	K_{jj'}(s) &= \frac{1}{2} \Biggl[ C_{jj'}\big((s+1)\Delta t\big) \notag \\
	&\quad - (s+1)K_{jj'}(0) - 2\sum_{h=1}^{s-1} (s+1 - h) K_{jj'}(h) \Biggr]\,,
\end{align}
using
\be
\label{K0}
K_{jj'}(0) = C_{jj'}(\Delta t)
\ee
as the initial value.
\Fig{Kjj'_compute} illustrates how \Eq{Cjj'} can be used to express cumulant elements $K_{jj'}(s)$ as linear combinations of particular values of the cumulant function $C_{jj'}(t)$.  For example, the cumulant element $K_{jj'} (1)$ can be computed using
\be
C_{jj'}(2\Delta t)=2K_{jj'}(0)+2K_{jj'}(1)\,.
\label{cumfig}
\ee
One further note is that the cumulant elements $K_{12}(s)$ require contributions from $K_{12}(s=0)$, meaning that at time steps $n$ and $m$, the system excitation is in both quantum dots. Physically, this never happens, but is introduced for mathematical consistency.

The linear polarization \Eq{PolarizationMthenWForster} then takes the form
\begin{align}
	\label{PolarizationKinim}
	P_{jk}(t) &= \sum_{i_{N-1} } \dots \sum_{i_{1} } M_{i_N i_{N-1}}
	\dots M_{i_1 i_{0}} \notag \\
	&\quad \times \exp\left(\sum_{n=1}^N \sum_{m=1}^N  \mathcal{K}_{i_n i_m}(s)\right).
\end{align}
A particular realization or a path in the system evolution is indicated by the indices $i_{N-1},\ldots, i_2,i_1 $. However, to obtain the full quantum dynamics of the system, all possible realizations are to be summed over, meaning a summation over all of these indices, which is done in \Eq{PolarizationKinim}. This is equivalent to the idea of path-integral approach.

\subsection{Tensor multiplication scheme}
\label{TensorScheme}
The memory time of a system under consideration is discretized due to the application of Trotter's decomposition, and the number of discrete time steps within this memory time corresponds to the temporal correlations which are taken into account. The $L$-neighbor approach is then used to describe the temporal correlations between all considered steps within the memory time. The cumulant elements $K_{jj'}(s)$, with $s=|n-m|$ increasing up to its maximum value, $L$, are found recursively from Eq.\eqref{recursiveK}.

By combining the contributions from matrices $\hat{M}$ and the exponential term in \Eq{PolarizationKinim} describing the exciton-phonon interactions across each time step,
we define a quantity $F^{(s)}$ which is generated via the recursive relation
\begin{equation}
	\label{Fn_app}
	F_{p i_L\ldots i_2}^{(s+1)} = \sum_{i_1=1,2,C} \mathcal{G}_{p i_L\ldots i_1} F_{i_{L}\ldots i_1}^{(s)}\,,
\end{equation}
using $F_{i_L\ldots i_1}^{(1)}= M_{i_1 k}$ as the initial value, which is equivalent to \Eq{Fn}, and $\mathcal{G}$ is the propagator given by \Eq{GtensorLN_mt}.

Using the cumulant elements \Eqs{Kinim}{recursiveK} allows us to provide an explicit expression for the propagator \Eq{GtensorLN_mt} for {\em Case 2}  :
\begin{align}
	\mathcal{G}_{p i_L \dots i_2 i_1} &= M_{i_2 i_1} \exp\left\{ \xi_{i_1}\xi_{i_1}K_{11}(0) + \eta_{i_1}\eta_{i_1}K_{22}(0) \right. \notag \\
	&\quad + 2\left(\xi_{i_2}\xi_{i_1}K_{11}(1) + \eta_{i_2}\eta_{i_1}K_{22}(1)\right) \notag \\
	&\quad + 2(\xi_{i_2}\eta_{i_1} + \eta_{i_2}\xi_{i_1})K_{12}(1) \notag \\
	&\quad + \dots \notag \\
	&\quad + 2\left(\xi_{p}\xi_{i_1}K_{11}(L) + \eta_{p}\eta_{i_1}K_{22}(L)\right) \notag \\
	&\quad + 2(\xi_{p}\eta_{i_1} + \eta_{p}\xi_{i_1})K_{12}(L) \left. \right\} \label{Gtensor_explicit_case2}
\end{align}
with each element of the tensor corresponding to a particular realization of the system.
\Eq{Gtensor_explicit_case2} reduces to a simpler form in {\em Case 1} since there is no second QD and only the single QD is interacting with phonons. It is given by
\begin{align}
	\mathcal{G}_{p i_L \dots i_2 i_1} &= M_{i_2 i_1} \exp\left\{ \xi_{i_1}\xi_{i_1}K_{11}(0) \right. \notag +2 \xi_{i_2}\xi_{i_1}K_{11}(1) \notag \\
	&\quad + \dots  + 2 \xi_{p}\xi_{i_1}K_{11}(L) \left. \right\}\, ,
	\label{Gtensor_explicit_case1}
\end{align}
with $i_n$ taking the values 0 or 1.
%\begin{align}
%	\mathcal{G}_{p i_L \dots i_2 i_1} &= M_{i_2 i_1} \exp\left\{ \xi_{i_1}\xi_{i_1}K_{11}(0) 2 \xi_{i_2}\xi_{i_1}K_{11}(1)  + \dots + 2 \xi_{p}\xi_{i_1}K_{11}(L)  \right\}\, .
%	\label{Gtensor_explicit_case1}
%\end{align}
\section{Real and virtual phonon-assisted transitions}\label{App:FGR}

FGR describes well the dephasing rates due to real phonon-assisted transitions. In the QD-cavity system with only two hybridized polariton states $\ket{\pm}$, separated by the Rabi splitting $R$, the dephasing rates calculated via FGR are given by $\Gamma_+(R)$ and $\Gamma_-(R)$, respectively, for downward and upward transitions, see the inset in \Fig{QDCAVITY_g600}. Here
\be
\Gamma_-(\omega) =  {\cal N}_{\omega} \,\Gamma_\text{\rm ph}(\omega)\,,\quad
\Gamma_+(\omega) =  ({\cal N}_{\omega}+1)\, \Gamma_\text{\rm ph}(\omega)\,,
\label{Eq:FGR}
\ee
where ${\cal N}_{\omega}$ is the Bose function, and
\begin{equation}\label{FGR_gamma_case1}
	\Gamma_\text{\rm ph}(\omega) = \frac{\pi}{4} \sum_{\bf q}\left| \lambda_{{\bf q}, 1}\right|^2 \delta( v_s q-\omega)\,,
\end{equation}
calculated below for a spherical QD for zero detuning.
The delta function in the above equation expresses the energy conservation in real transitions, indicating that the energy difference between the hybridized states should exactly match the energy of an emitted or absorbed phonon $\omega_q=v_s q$.

At large coupling strengths, the dephasing rate of state $\ket{+}$ is dominated by virtual phonon-assisted transitions involving the state $\ket{-}$, and vice versa.
Thus, a refinement of the above FGR result used in this work is the inclusion of virtual phonon-assisted transitions to second order \cite{muljarovDephasingQuantumDots2004}, so the dephasing rates of states $\ket{\pm}$ due to real and virtual transitions are given by
\begin{equation}\label{fulldephasingrates}
	\bar{\Gamma}_\pm (R) = \Gamma_\pm (R) + \frac{4}{R^2} \int_0^\infty \frac{d\omega}{\pi} \, \Gamma_+ (\omega) \, \Gamma_- (\omega)\, ,
\end{equation}
where $\Gamma_\pm (\omega)$ are defined in \Eq{Eq:FGR}. The second term in \Eq{fulldephasingrates} is therefore proportional to the fourth power of the exciton-phonon coupling matrix element and inversely proportional to $R^2$, where $R$ is the energy splitting between the states $\ket{+}$ and $\ket{-}$. This scaling follows from the unitary transformation of the polariton Hamiltonian, where the phonon-induced coupling between these states is eliminated to second order in perturbation theory \cite{muljarovDephasingQuantumDots2004}.

For an isotropic QD \cite{morreauPhononinducedDephasingQuantumdotcavity2019} in {\em Case 1},
$\Gamma_\textrm{ph}(R)$ is calculated by integrating over the phonon degrees of freedom, and \Eq{FGR_gamma_case1} is further expressed as
\begin{equation}
	\Gamma_\textrm{ph}(R) = R^3 \frac{(D_c - D_v)^2}{16\pi \rho_m v_s^5} \exp{-\frac{R^2l^2}{v_s^2}},
\end{equation}
with all the parameters defined in \App{App:Coupling}.

For {\em Case 2} with zero detuning, $\Gamma_\textrm{ph}(\omega)$ takes the form
\begin{equation}\label{FGR_gamma_case2}
	\Gamma_\text{\rm ph}(\omega) = \frac{\pi}{4} \sum_{\bf q}\left|\left(\lambda_{{\bf q}, 1}-\lambda_{{\bf q}, 2}\right)\right|^2 \delta( v_s q-\omega)\,,
\end{equation}
where the rate $\Gamma_\text{\rm ph}(R)$, evaluated in \cite{hallControllingDephasingCoupled2025}, provides for an isotropic model of the QDs the explicit analytical result:
\be
\Gamma_\text{\rm ph}(R) =\Gamma_0\left( 1-\frac{\sin(Rd / v_s)}{Rd /v_s } \right)\,,
\label{FGR}
\ee
where $\Gamma_0=R^3 (D_c - D_v)^2/(8\pi \rho_m v_s^5) e^{-l^2 R^2/v_s^2}$.


%\bibliography{bibliography}
%\bibliography{../../bibliography}
\end{document}